\documentclass[10pt,conference]{IEEEtran}
\IEEEoverridecommandlockouts
\usepackage{cite}
\usepackage{amsmath,amssymb,amsfonts}
\usepackage{graphicx}
\usepackage{textcomp}
\usepackage{xcolor}
\usepackage{algpseudocode}
\usepackage{booktabs}
\usepackage{quantikz}
\usepackage{pgfplotstable}
\newcommand{\codefont}[1]{\texttt{#1}}
\def\BibTeX{{\rm B\kern-.05em{\sc i\kern-.025em b}\kern-.08em
    T\kern-.1667em\lower.7ex\hbox{E}\kern-.125emX}}
\usepackage{graphicx}
\usepackage{caption}
\usepackage{subcaption}
\usepackage{textcomp}
\usepackage{amsmath,amsthm}
\usepackage{subcaption}
\usetikzlibrary{pgfplots.groupplots}
\usepackage{xcolor}
\usepackage[labelfont=bf]{caption}
\usepackage[linesnumbered,ruled]{algorithm2e}
\usepackage{amssymb}
\usepackage{colortbl}
\usepackage{tcolorbox}
\usepackage{listings}
\usepackage{makecell}
\usepackage{flushend}
\usepackage{xcolor}
\usepackage{tikz}
\usepgfplotslibrary{statistics}
\usepackage{enumitem}
\usepackage{multirow}
\usepackage[export]{adjustbox}
\usepackage{subcaption}

\definecolor{gray}{rgb}{0.5,0.5,0.5}
\lstdefinestyle{MyCSmallStyle} {
  language=Python,
  frame=none,
  xleftmargin=15pt,
  stepnumber=1, 
  numbers=left, 
  numbersep=5pt,
  numberstyle=\tiny\color[gray]{0.177}, 
  belowcaptionskip=\bigskipamount,
  captionpos=b, 
  escapeinside={*'}{'*},
  tabsize=5,
  emphstyle={\bf},
  escapechar=!,
  basicstyle=\scriptsize\ttfamily,
  keywordstyle=\color{javapurple}\bfseries,
  stringstyle=\color{javared},
  commentstyle=\color{javagreen},
  morecomment=[s][\color{javadocblue}]{/**}{*/},
  showspaces=false,
  columns=flexible,
  showstringspaces=false,
  morecomment=[l]{//},
  tabsize=2,
  breaklines=true,
  moredelim=[is][\underbar]{^}{^}
}

\definecolor{darkBlue}{rgb}{0.000000,0.000000,0.545098}
\definecolor{darkGreen}{rgb}{0.000000,0.392157,0.000000}
\definecolor{DarkGray}{gray}{0.4}
\definecolor{javared}{rgb}{0.6,0,0} 
\definecolor{javagreen}{rgb}{0.25,0.5,0.35} 
\definecolor{javapurple}{rgb}{0.5,0,0.35} 
\definecolor{javadocblue}{rgb}{0.25,0.35,0.75} 
\definecolor{lightgray}{gray}{0.95}
\definecolor{shadecolor}{RGB}{150,150,150}
\definecolor{blueA}{RGB}{204,229,255}
\definecolor{redA}{RGB}{112,0, 0}

\newcommand{\eg}{\hbox{\emph{e.g.}}\xspace}
\newcommand{\ie}{\hbox{\emph{i.e.}}\xspace}

\newcommand{\codeIn}[1]{{\small\texttt{#1}}}

\newcommand{\tool}{{\small{\textsc{QDiff}}}\xspace}
\newcommand\jiyuan[1]{\textcolor{purple}{JW: #1}}

\newtheorem{definition}{Definition}[section]

\newcommand{\MyPara}[1]{\vspace{.3em}\noindent\textit{\textbf{#1}}}

\begin{document}

\title{QDiff: Differential Testing of \\Quantum Software Stacks}

\author{
\IEEEauthorblockN{Jiyuan Wang, Qian Zhang, Guoqing Harry Xu, Miryung Kim}
\IEEEauthorblockA{
\textit{University of California, Los Angeles}\\
CA, United States \\
\{wangjiyuan, zhangqian, harryxu, miryung\}@cs.ucla.edu}

}

\maketitle

\begin{abstract}
Over the past few years, several quantum software stacks (QSS) have been developed in response to rapid hardware advances in quantum computing. A QSS includes a quantum programming language, an optimizing compiler that translates a quantum algorithm written in a high-level language into quantum gate instructions, a quantum simulator that emulates these instructions on a classical device, and a software controller that sends analog signals to a very expensive quantum hardware based on quantum circuits.
In comparison to traditional compilers and architecture simulators, QSSes are difficult to tests due to the probabilistic nature of results, the lack of clear hardware specifications, and quantum programming complexity. 

This work devises a novel \emph{differential testing} approach for QSSes, named \tool with three major innovations:  (1) We generate input programs to be tested via  semantics-preserving, source to source transformation to explore program variants. (2) We speed up differential testing by filtering out quantum circuits that are not worthwhile to execute on quantum hardware by analyzing static characteristics such as a circuit depth, 2-gate operations, gate error rates, and T1 relaxation time. (3) We design an extensible equivalence checking mechanism via distribution comparison functions such as Kolmogorov–Smirnov test and cross entropy.

We evaluate \tool with three widely-used open source QSSes: Qiskit from IBM, Cirq from Google, and Pyquil from Rigetti. By running \tool on both real hardware and quantum simulators, we found several critical bugs revealing potential instabilities in these platforms. \tool's source transformation is effective in producing semantically equivalent yet not-identical circuits (i.e., 34\% of trials), and its filtering mechanism can speed up differential testing by 66\%.

\end{abstract}


\section{Introduction}
\label{sec:intro}

Quantum computing is an emerging computing paradigm that promises unique advantages over classical computing. However, quantum programming is challenging. Quantum computation logic is expressed in qubits and quantum gates, and the states of quantum registers are measured probabilistically. Due to the physical properties of quantum mechanics, qubits and quantum gates are fundamentally different from classical bits and gate logic. For example, \emph{quantum indeterminacy} dictates that the same quantum program can produce different results in different executions. 

To this end, many quantum software stacks such as Google's Cirq~\cite{Cirq}, Rigetti's  Pyquil~\cite{pyquil_paper}, and IBM's  Qiskit~\cite{qiskit_rpo} have been developed to provide user-friendly high-level languages for quantum programming, abstracting away the underlying physical and mathematical intricacies. 

A quantum software track (QSS) includes (1) APIs and language constructs to express quantum algorithms, (2) a compiler that transforms and optimizes a given input quantum algorithm at the circuit level, and (3) a backend executor that either simulates the resulting gates on classical devices or executes directly on quantum hardware. Consider Qiskit~\cite{aleksandrowicz2019qiskit}, which consists of four components: \codeIn{QiskitTerra}~that compiles and optimizes programs, \codeIn{QiskitAer}~that supports high-performance noisy simulations, \codeIn{QiskitIgnis} for error correction, noise characterization, and hardware verification, and \codeIn{QiskitAqua} that helps a developer to express a quantum algorithm or an application. Other QSSes~\cite{pyquil_paper} such as Pyquil have a similar set of components\textemdash \eg, Pyquil uses \codeIn{quilc} as a compiler and \codeIn{qvm} as a quantum simulator. 

As with any compiler framework, a QSS could be error-prone. Developers and users often report bugs on popular QSSes~\cite{pyquil_issue,Cirq_issue,qiskit_rpo}, and a simple search on StackOverflow with the keyword ``quantum error'' would bring up over 500 posts on various QSS components, ranging from compiler settings, simulation, and the actual hardware~\cite{stackoverflow}. These posts often reveal deeper confusion that developers face due to the inherent probabilistic nature of quantum measurements\textemdash {\em if a program produces a result that looks different from what is expected, is it due to a bug or the non-determinism inherent in quantum programs? Is there divergence beyond expected noise coming from an input program, a compiler, a simulator, and/or hardware?}

\MyPara{Challenges.}
There are three technical challenges that make it difficult to test QSSes. 

The first challenge is how to generate semantically equivalent programs for testing quantum compilers~\cite{amy2014polynomial,nam2018automated}. Though existing compiler testing techniques generate equivalent programs by manipulating code in dead regions or unexecuted branches~\cite{le2014compiler}, a quantum program usually does not have many unexecuted branches. Therefore, we must design a technique that can produce a large number of semantically-equivalent variant circuits that may induce divergence on hardware. 

The second challenge is that compilers are not the only source of bugs in a QSS. For example, in a line of recent work~\cite{hietala2019verified, shi2019contract} on verified quantum compilers, compilers are guaranteed to be correct with respect to their optimization and transformation rules. However, the overall correctness of QSSes goes beyond compiler optimization rules. We must reason about subsequent execution on quantum simulators and hardware, which are often the major sources of instabilities. In fact, Rigetti or Cirq's Github histories indicate that bugs may come from its various backends due to improper initialization~\cite{pyquil_error}, problematic APIs~\cite{apiserror}, or inappropriate synthesis options~\cite{Cirq_error}.  

The third challenge is how to interpret measurements from quantum simulators or hardware, since quantum measurement operations by definition are probabilistic in nature due to quantum indeterminacy~\cite{hietala2019verified}. Therefore, when we execute logically equivalent program variants, comparing their measurement results is not straightforward, as we must take into account inherent noise (e.g., hardware gate errors, readout errors, and decoherence errors) and must determine whether the observed divergence is significant enough to be considered as a meaningful instability.



\MyPara{This Work.} \tool is a novel differential testing technique for QSSes. \tool takes as input a seed quantum program and reports potential bugs with a pair of \emph{witness programs} (i.e., logically equivalent programs that are supposed to produce identical behavior) but triggers divergent behavior on the target QSS (i.e., meaningful instabilities beyond expected noise). 
\tool overcomes the aforementioned challenges with three technical innovations below. 

First, \tool generates logically equivalent quantum programs using a set of \emph{equivalent gate transformation} rules. This is based on the insight that each quantum gate can be represented mathematically as a unitary matrix; a sequence of gates essentially corresponds to the multiplication of their unitary matrices, which also yields a unitary matrix. As such, one sequence of gates is semantically equivalent to another if the two sequences yield the same unitary matrix. We leverage seven gate transformation rules to generate different gate sequences that are guaranteed to yield the same unitary matrix.
These sequences essentially define logically equivalent program variants that are supposed to yield identical behavior. Such variants are then executed on a simulator or hardware and their measurements are compared. 

Second, with multiple logically equivalent variants generated after optimization, \tool selects a subset of those circuits that are worthwhile to run on a noisy simulator or quantum hardware. 
This selection and filtering process is crucial for both {\em speedup} and {\em potential cost saving}, because there are only a very few quantum hardware devices in the world, and they are {extremely expensive} resources\textemdash D-wave's quantum computer is reported to be \$15 million in 2017~\cite{Dwave}. IBM Qiskit allows public hardware access up to 5 qubits, but a higher number of qubit usage is strictly restricted. Noisy simulators are extremely compute-intensive to run on classical computers as well, taking on more than 4 minutes to do 3000 measurements on a simple four-qubit circuit. This noisy simulation cost becomes extremely unmanageable with the higher number of qubits.  

To achieve speed up, \tool analyzes the static characteristic of logically equivalent circuits such as the circuit depth (i.e., {\em moments}), the number of 2-qubit gates, the known error rates of each gate kind, and T1 relaxation time for quantum hardware. \tool obviates the need of executing certain circuits on hardware or noisy simulators, if the logically equivalent circuits are incapable of revealing {\em meaningful instabilities} in QSS. 

Third, in order to compare executions of logically equivalent circuits, \tool first identifies how many measurements are needed for reliable comparison. Based on the literature of distribution sampling, \tool adapts closeness testing in $L1$ norm~\cite{chan2014optimal} to estimate the required number of measurements given confidence level $p=2/3$. Using either a noisy simulator or directly on hardware, it then performs the required number of measurements. To compare two sets of measurements, \tool uses distribution comparison methods. Two methods based on Kolmogorov-Smirnov (K-S) test~\cite{kolmogorov1933sulla,smirnov1939estimation} and Cross~Entropy~\cite{good1956some,shore1980axiomatic} are currently supported. This comparison method is easily extensible in \tool, as a user can define a new statistical comparison method~\cite{chan2014optimal} in \tool. 


\MyPara{Results.} We evaluated \tool with the latest versions of three widely-used QSSes: Qiskit, Cirq, and Pyquil. With six seed quantum algorithms, \tool generates 730 variant algorithms through semantics-modifying mutations. Starting from the generated algorithms, it generates a total of 14799 program variants using semantics-preserving source transformations. This generation process took {\tool} 14 hours. With the filtering mechanism, \tool reduces its testing time by 66\%. Using \tool, we determined total 6 sources of instabilities. These include 4 software crash bugs in Pyquil and Cirq simulation, and 2 potential root causes that may explain 25 out of 29 cases of divergence beyond expected noise on IBM hardware.


The main contributions of this work include:

\begin{itemize}
\item We present the first differential testing framework for QSSes. This framework is integrated with three widely used QSSes from Google, IBM, and Rigetti.  
\item We embody three technical innovations in \tool to make differential testing fast and to increase the chance of finding meaningful instabilities beyond the usual expected noise: (1) auto-generation of logically equivalent variants, (2) auto-filtering of circuits that are worthwhile to run on hardware (or a noisy simulator), and (3) pluggable methods of comparing measurements in terms of statistical distribution. 
\tool obviates the need of invoking quantum back-ends by 66\%. Using \tool, we determined 6 sources of instabilities in three widely used QSSes: 4 crash bugs and 2 root causes for divergences beyond expected noise on IBM hardware.   
\end{itemize}

We provide access to the artifacts of \tool at 
https://github.com/wjy99-c/QDiff.

\section{Background}
\label{sec:back}

\begin{table*}[t]
\centering
\caption{Different layers that bugs appear}
\vspace{-2ex}
\label{tab:errortypes}
\scalebox{1}{\begin{tabular}{lrl}
\toprule
\multicolumn{1}{c}{\bf Layers} & \multicolumn{1}{l}{\bf Percentage} & \multicolumn{1}{c}{\bf Example}  \\ \hline

Compiler (optimizations \& settings) & 53.9\% & \codeIn{MergeInteractions()} returns different results for the same circuit \cite{investigate_API_error}.  \\  \hline

Backend (simulators \& hardware) & 19.7\% & Simulators change a global random state on Cirq \cite{investigate_backend_error}.\\ \hline

API and quantum gate & 11.8\% & PhasedXPowGate raised to a symbol power fails the \codefont{is\_parameterized} protocol~\cite{investigate_gate_error}.  \\

\bottomrule
\end{tabular}}
\vspace{-1mm}
\end{table*}

\MyPara{Quantum Computing.}
Quantum computing emerges as a promising technology for many domains where quantum computers have been demonstrated to outperform classical computers by a large margin. For example, Grover's algorithm~\cite{grover1996fast} can find a given item in an unsorted database with a $\sqrt{N}$-times speedup when running on a quantum computer, compared to a classical computer. Similar to programs executed on heterogeneous devices such as FPGAs, a quantum application is comprised of {\em host code} that runs on CPU and {\em quantum code} that runs on quantum hardware. Generally, quantum computers work as accelerators to execute the compute-intensive parts of the original application. For example, the Shor algorithm~\cite{shor1994algorithms} can achieve an exponential speedup by decomposing integer factorization into a reduction to be executed on a classical computer and an order finding problem to be executed on quantum hardware.

\MyPara{Quantum Software Stack.}
Currently, three most widely used QSSes are Qiskit, Cirq, and Pyquil~\cite{larose2019overview}.

\begin{itemize}
\item{\bf Qiskit~\cite{qiskit_rpo}} is an open-source framework developed by IBM. Qiskit provides a software stack that is easy to use quantum computers and facilitates quantum computing research. Qiskit consists of \codeIn{Qiskit Terra} (compiler), \codeIn{Qiskit Aer} (several quantum simulators), \codeIn{Qiskit Ignis} (which supports error correction and noise characterization), and \codeIn{Qiskit Aqua} (APIs to help developers write applications).

\item{\bf Cirq~\cite{Cirq_rpo}} is an open-source Python framework from Google. It enables a developer to create and simulate Noisy Intermediate-Scale Quantum (NISQ) circuits. Cirq consists of an \codeIn{optimization} component to compile and transform circuits, a \codeIn{simulator} component for emulating quantum computation on classical devices, and other \codeIn{development} libraries for application development.

\item{\bf Pyquil~\cite{pyquil_rpo}} is an open-source Python framework developed by Rigetti. It builds on Quil, an open quantum instruction language for near-term quantum computers, and uses a combined classical/quantum memory model. PyQuil is the main component of Forest, the overarching platform for Rigetti’s quantum software. Pyquil consists of: \codeIn{quilc} (compiler), \codeIn{qvm} (quantum virtual machine with several kinds of simulators), and \codeIn{pyquil} (a library to help users write and run quantum applications). 

\end{itemize}

All three QSSes are similar to one another in that each includes a quantum programming language, an optimizing compiler that outputs quantum gate instructions, a quantum simulator that emulates these instructions on a classical device, and a software controller that executes gate instructions on quantum hardware.

\MyPara{Quantum Bit.} A {\em quantum bit}, or {\em qubit} for short, is the basic unit of quantum computation.
Unlike a classical bit that is either $0$ or $1$, a qubit's state is a probabilistic function of $|0\rangle$ and $|1\rangle$, represented as:
\begin{equation}
    |\psi\rangle=\alpha  |0\rangle+\beta |1\rangle, \text{where}  \left| \alpha \right| ^{2}+\left| \beta \right| ^{2}=1
\end{equation}
The state of a qubit is unknown until a \emph{measurement} is completed, resulting in $\left| \alpha \right| ^{2}$ to be 0 and $\left| \beta \right| ^{2}$ to be 1.
Naturally, the sum of the probabilities (i.e., the modulus squared of amplitudes) is 1: $\left| \alpha \right|^{2}+\left| \beta \right|^{2}=1$. 


\MyPara{Quantum Gate.}
Classical computers use logic gates to transform signals from input wires. For example, a \textit{NOT} gate, also known as an inverter, takes a single signal as input and outputs the opposite value. The quantum gate analogous to \textit{NOT} is an \textit{X} gate, which transforms a qubit $\alpha |0\rangle+\beta |1\rangle$ to $\beta|0\rangle+\alpha |1\rangle$.

An \textit{X} gate has the following matrix-based representation:
\begin{equation}
    X=\begin{bmatrix} 0 & 1 \\ 1 & 0 \end{bmatrix}
\end{equation}
Using a vector to represent the quantum state $\alpha|0\rangle+\beta |1\rangle$,
applying an \textit{X} gate has the following effect:
\begin{equation}
    X\begin{bmatrix} \alpha  \\ \beta  \end{bmatrix}=\begin{bmatrix} \beta  \\ \alpha  \end{bmatrix}=\begin{bmatrix} \alpha'  \\ \beta' \end{bmatrix}
\end{equation}
Other commonly-used quantum gates include $H$, $T$, $CNOT$, $Z$, $\mathit{CZ}$, $\mathit{S}$, $\mathit{U1}$, and $\mathit{U3}$; a full explanation of these gates is elsewhere~\cite{nielsen2002quantum}. 

Since all quantum gates can be represented as \emph{matrices}, we can transform a sequence of gates to another logically equivalent sequence without altering its outcome, as long as \emph{the multiplication of the matrices for gates in each sequence} produces the same result. As an example, a \textit{SWAP} gate, a two-qubit gate that swaps the two qubits' states, is semantically equivalent to a sequence of three \textit{CNOT} gates.
\begin{equation}\small
    \textit{SWAP}(q_{1},q_{2})=\textit{CNOT}(q_{1},q_{2})\textit{CNOT}(q_{2},q_{1})\textit{CNOT}(q_{1},q_{2})
\end{equation}

This observation forms the foundation of \tool's program variant generation procedure, detailed in Section~\ref{sec:appr}.

\MyPara{Quantum Circuit.} A quantum circuit consists of a set of connected quantum gates. Since quantum gates can be represented as unitary matrices, a quantum circuit is essentially the multiplication of the matrices. 

Figure~\ref{fig:circuit_code} shows a program in IBM's Qiskit~\cite{qiskit_rpo}. It first registers two qubits and initializes them to $|0\rangle$. Next, an \textit{H} gate sets the first qubit into state $\frac{1}{\sqrt{2}}|0\rangle + \frac{1}{\sqrt{2}}|1\rangle$ and an \textit{X} gate flips the second qubit from $|0\rangle$ to $|1\rangle$. Finally, a measurement is performed and stored in a classical array. The function returns this circuit as a function-type value.

\begin{figure}[t]
        \begin{subfigure}[t]{1\linewidth}
        \input{code/background_circuit}
        \vspace{-1ex}
        \caption{A quantum circuit with an $H$ gate and an $X$ gate.}
            \label{fig:circuit_code}
        \end{subfigure}
        \begin{subfigure}[t]{1\linewidth}

        \input{code/background_program}
        \caption{The host code using the circuit in (a).}
            \label{fig:program_code}
        \end{subfigure}
      \vspace{-0.5em}
    \caption{An example Qiskit program.
    \label{fig:quantum_example}}
    \vspace{-2ex}
\end{figure}
Figure~\ref{fig:program_code} shows host code that calls this quantum circuit. Users can execute this circuit on different backends such as real quantum hardware or simulators. The circuit is executed on \textit{qasm\_simulator} for a thousand times (\codeIn{shots=1000}). Since the first qubit state is $\frac{1}{\sqrt{2}}|0\rangle + \frac{1}{\sqrt{2}}|1\rangle$ and the second qubit state is $|1\rangle$, each run produces a result of either $|01\rangle$ or $|11\rangle$ with the equal probability of $0.5 =(\frac{1}{\sqrt{2}})^{2}$.

\MyPara{T1 relaxation time.} 
In quantum computing, a qubit can retain data for only a limited amount of time, referred to as \emph{relaxation Time} because a qubit in a high-energy state (state $|1\rangle$) naturally decays to a low-energy state (state $|0\rangle$). The time span for this decay is referred to as {\em T1 Relaxation Time}. For a physical circuit, its measurement results are unreliable if its execution time is longer than {\em T1}.


\section{Motivation}
\label{sec:motivation}
To understand real-world QSS bugs, we collected 76 latest issues reported on Github Pyquil, Cirq, and Qiskit. After excluding 11 issues related to installation and other tools, we categorized the remaining 65 issues by the layer where each issue appears: compilers, backends, and APIs.

Table~\ref{tab:errortypes} summarizes the percentage of the corresponding layer and a representative example. Most bugs appear at the compiler level---compiler optimizations and settings. For example, Cirq produces incorrect circuits when using \codeIn{MergeInteractions()}~\cite{investigate_API_error}. The second most common bugs appear at the backend level\textemdash simulators and hardware execution, \eg, the initial state is unexpectedly modified by the Cirq simulator in~\cite{investigate_backend_error}. The other issues are with respect to the API implementation of high-level gates. For example, in Cirq, when \codeIn{cirq.PhasedXPowGate} is invoked with an input argument \codeIn{exponent}, invoking \codeIn{is\_parameterized} on the resulting gate should return \codeIn{true} but returns \codeIn{false} due to a bug in \codeIn{cirq.PhasedXPowGate}.

\begin{figure}[t]
    \centering
    \scalebox{0.85}{
    \includegraphics[width=\linewidth]{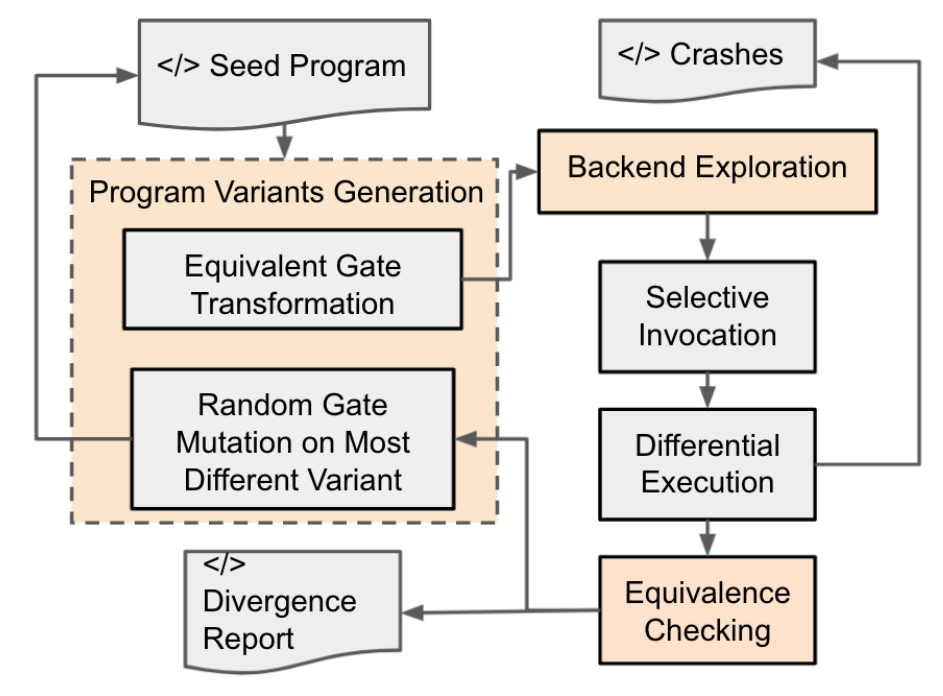}}
    \caption{\tool Overview}
    \vspace{-2em}
    \label{fig:workflow}
\end{figure}

The aforementioned bugs are hard to find due to quantum indeterminacy. The following excerpt illustrates a concrete example of how one google tutorial user is confused whether a problem is a bug or due to inherent non-determinism. Cirq's VQE (Variational-Quantum-Eigensolver) tutorial had a mistake\textemdash the orders of mix layer and cost layer were described in a wrong order. The tutorial user \codeIn{D} (shortened) posted a question on StackExchange~\cite{cirq_compilerbugs} with the embedded code from Cirq's VQE tutorial. A Cirq developer \codeIn{C} (shortened) noticed that ``The strange thing is the example output shows the output probabilities varying with $\mathit{gamma}$ (the CZ parameter)", where $\mathit{gamma}$ should not have any effect on the measurement results.'' \codeIn{D} actually believed the disagreement is due to quantum indeterminacy, until \codeIn{C} explicitly labeled it as a compiler bug. 

\begin{table*}[t]
    \centering \caption{Explored Compiler Configurations}
    \label{tab:pragma}
    \vspace{-5pt}
    \scalebox{0.97}{\scalebox{0.9}{\begin{tabular}{c|l|l}
\toprule
    {\bf Framework} & {\bf Options} &\makecell[c] {\bf Description}\\\hline

{} & \codeIn{BasicSwap, LookaheadSwap,} & Specify how swaps should be inserted to make the circuit compatible with the \\
& \codeIn{StochasticSwap} & coupling map. \tool checks if BasicSwap has the most $\mathit{SWAP}$ gates. \\\cline{2-3}

Qiskit  & \codeIn{Optimization$\_$level=0,1,2,3} & Specify the optimization level\textemdash the higher the level, the simpler the resulting circuit.\\
& & \tool checks if a higher-level optimization generates a more complex circuit.\\\hline
        
& \codeIn{DropEmptyMoment()}
& Remove empty moments from a circuit. \tool checks empty moments in the circuit.\\\cline{2-3}

Cirq & \codeIn{MergeInteractions()} & Merge adjacent gates; \tool checks the applicability of G6 in Table~\ref{tab:trans_rules}.\\\cline{2-3}

{}&\codeIn{PointOptimizationSummary()}& User-defined optimization. \\\hline
 & \codeIn{PRAGMA INITIAL\_REWIRING \{"NAIVE"} & Change the optimization/mapping style \\
& \codeIn{,"GREEDY", "PARITIAL"\}}& \\\cline{2-3}
Pyquil& \codeIn{PRAGMA \{COMMUTING\_BLOCKS} & Change the optimization style for certain part of the code.\\
&\codeIn{,PERSERVE\_BLOCKS\}} & \\

\bottomrule   
\end{tabular}}}
\end{table*}

\MyPara{Key Takeway.} This study of Github issues shows that QSS bugs are not just traditional compiler bugs and may come from various sources at different layers: API implementation of high-level gates, backend simulators, or hardware execution. This problem of detecting QSS bugs is further complicated by the probabilistic and noisy nature of quantum indeterminacy. 

\MyPara{}

\section{\tool Approach}
\label{sec:appr}

\tool contains three novel components to detect meaningful instabilities in QSSes. Figure~\ref{fig:workflow} shows {\em program variant generation} using equivalent gate transformation and mutations (Section~\ref{sec:variant});  {\em backend exploration} leveraging selective invocation of quantum simulators and hardware (Section~\ref{sec:exploration}); and {\em equivalence checking} via distribution comparison (Section~\ref{sec:equivalence_checking}). Starting from a seed program, \tool iterates these steps until a time limit is reached. Our key insight is that (1) we can generate semantically equivalent but syntactically different circuits, and (2) we can speed-up the differential execution process by filtering out certain circuits, because they will definitely lead to unreliable divergence and thus are {\em not worthwhile to run} on hardware or noisy simulators. 

\subsection{Program Variant Generation}\label{sec:variant}
Prior work~\cite{le2014compiler} finds that testing compilers with equivalent programs is highly effective. \tool adapts this idea to the domain of quantum compiler testing by creating logically equivalent gate sequences. It then checks whether the corresponding equivalent circuits produce the same results (in this case, a similar statistical distribution with some noise) on quantum simulators or hardware. For this purpose, \tool generates program variants by repeating the two-fold process of {\em applying semantics-preserving gate transformation} to each generated program in each iteration and {\em applying semantics-modifying mutations} to diversify the pool of input programs in the next iteration.

\MyPara{Equivalent Gate Transformation (EGT).}
\begin{table}[t]\small
    \centering    \caption{\tool generates program variants based on EGT rules G1-G7.
    \label{tab:trans_rules}}
    \scalebox{0.95}{
    \begin{tabular}{cllc}
    \toprule
        {\bf Rule ID} & {\bf Original Construct} & {\bf Equivalent Construct} \\\hline
        
        & & CNOT($q_{1}$,$q_{2}$)\\
        G1 & SWAP($q_{1}$,$q_{2}$) & CNOT($q_{2}$,$q_{1}$) \\
        & & CNOT($q_{1}$,$q_{2}$)\\\hline
        
        G2& H($q_{1}$)H($q_{1}$)  & Merged to Identity Matrix\\ \hline
        
        G3 & X($q_{1}$) & H($q_{1}$)S($q_{1}$)S($q_{1}$)H($q_{1}$) \\\hline
        
        G4 & Z($q_{1}$) & S($q_{1}$)S($q_{1}$) \\\hline
        
        G5 & CZ($q_{1}$,$q_{2}$) & H($q_{2}$)CNOT($q_{1}$.$q_{2}$)H($q_{2}$) \\\hline
        
        G6 & CZ($q_{1}$,$q_{2}$)CZ($q_{1}$,$q_{2}$)&  Merged to Identity Matrix \\\hline
        G7 & CCNOT($q_{1}$,$q_{2}$,$q_{3}$)&6 CNOT gates \\
        & & with 9 one-qubit gates\\

    \bottomrule
    \end{tabular}
    }
    \vspace{-4ex}
\end{table}

As discussed earlier in Section~\ref{sec:back}, one quantum gate sequence is semantically equivalent to another sequence, if they both yield the same unitary-matrix representation.
In fact, complex quantum gates, without altering the outcome, can be described as a combination of basic quantum gates. \tool leverages seven gate transformation rules to map complex gates into sequences of simple gates, as shown in Table~\ref{tab:trans_rules}. In G7, a Toffoli gate (\ie $CCNOT$) can be replaced with 6 $CNOT$ gates and 9 one-qubit gates. To explore the alternative representation of a given gate sequence, \tool finds the applicable transformation rules by matching the gate names. For an example circuit $S(q2)Z(q1)$, \tool identifies the complex gate $Z(q1)$, applies rule G2 to construct gate $Z(q1)$ as a sequence $S(q1)S(q1)$, and generates a final variant $S(q2)S(q1)S(q1)$. As opposed to an optimizing compiler that applies transformation rules to reduce the number gates or the total gate depth~\cite{hietala2019verified,qiskit_rpo}, \tool aims to diversify the pool of input programs through source to source transformation. 

\begin{table*}[t]
    \centering
    \caption{IBM quantum hardware's gate-level error rates, gate time, and T1 relaxation (decoherence) time.}
        \begin{tabular}{c|c|c|c|c|c}
    \hline
        {\bf IBM quantum computer} & {\bf T1 time /$\mu s$} & {\bf Gate time /$ns$} &{\bf 2-qubit gate error rate} & {$t_{m}$} &{$\delta_{2qubit}$} \\\hline
        ibm\_santiago & 155.19 & 408.89 & 1.79\% & 379 &5 \\
        ibm\_yorktown & 56.81 & 476.44 & 2.03\% &119 &5 \\
        ibm\_16\_melbourne & 53.37 & 928.71 & 3.29\% &57 &3 \\
        ibm\_belem & 74.45 & 552.89 & 1.07\% &135 &11 \\
        ibem\_quito & 85.35 & 353.78 & 1.03\% &241 &11 \\\hline
    \end{tabular}

    \label{tab:threshold_QHW}
\end{table*}

\MyPara{Seed Diversification via Mutations.}\label{sec:mutation} While generation of logically equivalent programs can find bugs through observing disagreements, they are unlikely to exercise various programming constructs. To diversify the seed input programs and to explore hard-to-reach corner cases in quantum compilers. \tool borrows the idea of mutation-based fuzz testing~\cite{afl} and designs a set of semantics-modifying mutations. 

After generating multiple logically equivalent programs in each iteration, \tool calculates the average distributions among the equivalent programs as the reference distribution, then picks the variant that leads to the largest comparison distance (e.g., K-S distance) from the reference distribution, and randomly applies one of the following four mutation operations to the variant in order to generate a new algorithm. This is based on the insight that the program with the most deviating results has a higher chance to expose unseen behavior~\cite{le2014compiler,lidbury2015many}. Please note that these mutations do not preserve semantics; instead, the goal of mutations is to resume the next round of differential testing with a different algorithm. We start with four mutation operators listed below, used in quantum mutant generation~\cite{mendiluzemuskit,ali2021assessing}.


\begin{itemize}
\item{\bf Gate Insertion/ Deletion (M1)} inserts/deletes random quantum gates: \eg, insert \fbox{\begin{minipage}[t][0.5em][t]{7em}\centering
\centerline{\texttt{\footnotesize{prog.x(qubit[1])}}}
\end{minipage}};
\item {\bf Gate Change (M2)} changes a quantum gate to another gate: \eg, from \fbox{\begin{minipage}[t][0.5em][t]{7em}\centering
\centerline{\texttt{\footnotesize{prog.x(qubit[1])}}} 
\end{minipage}} to \fbox{\begin{minipage}[t][0.5em][t]{7em}\centering
\centerline{\texttt{\footnotesize{prog.h(qubit[1])}}}
\end{minipage}};
\item {\bf Gate Swap (M3)} swaps two quantum gates;
\item {\bf Qubit Change (M4)} changes the qubits: \eg, from \fbox{\begin{minipage}[t][0.5em][t]{7em}\centering
\centerline{\texttt{\footnotesize{prog.x(qubit[1])}}} 
\end{minipage}} to \fbox{\begin{minipage}[t][0.5em][t]{7em}\centering
\centerline{\texttt{\footnotesize{prog.x(qubit[2])}}} 
\end{minipage}}.
\end{itemize}

\subsection{Quantum Simulation and Hardware Execution} \label{sec:exploration}

\MyPara{Compiler Configuration Exploration.}
\tool automatically explores different compiler configurations. In Qiskit, compiler settings can be specified by the arguments passed to the backends e.g., users can apply \codeIn{optimization level=1} to collapse adjacent gates via light-weight optimization, while \codeIn{optimization level=3} does heavy-weight optimization to resynthesize two-qubit blocks in the circuit. In Cirq, compiler settings must be specified using API invocations: e.g., users can write their own optimization with \codeIn{PointOptimizationSummary()}. In Pyquil, compiler settings are specified using inlined pragmas similar to how FPGA developers specify high level synthesis options using pre-processor directives, e.g., a region denoted by \codeIn{PRAGMA PRESERVE\_BLOCK} will not be modified by a compiler. There are in total 2, 3, and 2 configuration types for Qiskit, Cirq, and Pyquil respectively, as shown in Table~\ref{tab:pragma}. 
 When executing a variant with a specific compiler setting, \tool records both thrown exceptions and program timeouts.
 
\begin{table}[t]
    \centering \caption{Explored Backends (Simulators and Hardware)}
    \label{tab:backends}
    \vspace{-5pt}
    \scalebox{0.97}{\scalebox{0.9}{\begin{tabular}{c|l|l}
\toprule
{\bf Framework} & {\bf Backends} & {\bf Description}\\\hline

{} & \codeIn{statevector\_simulator} & noiseless sim. \\\cline{2-3}
& \codeIn{qasm\_simulator} & noiseless sim. \\\cline{2-3}

Qiskit  & \codeIn{FakeSantiago} & \\
& \codeIn{FakeYorktown} &noisy sim. \\
& \codeIn{FakeMelbourne} &
\\\cline{2-3}

& \codeIn{ibmq\_santiago} &\\
& \codeIn{{ibmq\_yorktown}} &quantum hardware\\
& \codeIn{ibmq\_16\_melbourne} &\\\hline

& \codeIn{Simulator}
& noiseless/noisy sim. \\\cline{2-3}

Cirq & \codeIn{DensityMatrixSimulator} & noiseless/noisy sim. \\\hline

&\codeIn{Aspen-x-yQ-noisy-qvm} & noisy sim. \\\cline{2-3}
Pyquil& \codeIn{WavefunctionSimulator} & noiseless sim.\\\cline{2-3}
&\codeIn{Aspen-x-yQ-qvm,Pyqvm} & noiseless sim. \\

\bottomrule   
\end{tabular}}}
    \vspace{-1em}
\end{table}
\MyPara{Backend Exploration.} Backend exploration runs the same input program on different backends, shown in Table~\ref{tab:backends}. In terms of real hardware execution, \tool uses the free version of IBM hardware only, because other platforms are currently proprietary. \tool is extensible by specifying a different backend configuration. Noisy simulators and state-vector simulators are both included in \tool's backend exploration.

\MyPara{Filtering and Selective Invocation on Hardware.} \tool is focused on isolating software defects, not hardware defects. Because hardware imperfections such as decoherence is present, it is important to filter out circuits that would invoke errors due to the inherent hardware limitations. With the above observations, \tool filters out unnecessary circuits in two steps. First, \tool examines the final gate sequences after all compiler optimizations and logical-to-physical mappings, filtering out exactly identical physical circuits by moment-by-moment comparison. Second, \tool analyzes the static characteristics of circuits to remove those that certainly produce unreliable executions (i.e. results dominated by hardware-level noise such as gate errors and relaxation errors and hence unreliable), while leaving those that may produce {\em meaningful divergences} using Definition IV.1. 

As discussed in Section~\ref{sec:back}, for a physical circuit, its measurement results are unreliable if its execution time is longer than T1, implying that the number of circuit moments $n_{m}$ (the depth of the circuit) in any circuit should not exceed a threshold $t_{m}$. Moreover, different kinds of quantum gates have different inherent error rates. For publicly available IBM quantum computers, error rates of single-qubit operations are in the order of 10$^{-3}$, while error rates of 2-qubit gate operations are in the order of 10$^{-2}$. A typical quantum program contains a significant number of 2-qubit gates, whose errors contribute the most to the overall error rate because such gates are more error-prone than 1-qubit gates~\cite{tannu2019not}. Taking this into consideration, $d_{2qubit}$ (the difference in the number of 2-qubit gates from the original circuit) should not exceed an application-specific threshold $\delta_{2qubit}$ to avoid unreliable results. Leveraging the above observations, we define the {\em worthiness} of invoking a quantum circuit.

\begin{definition}
A circuit is worth invoking on quantum hardware or noisy simulator, if it satisfies the following condition: $n_{m}<t_{m}$ and $d_{2qubit}<\delta_{2qubit}$.
\label{th_def}
\end{definition}

The threshold $t_{m}$ is determined empirically by two factors: (1) IBM computers' average $T1$ time, and (2) the average $\mathit{gate}$ execution time for all gates, as listed in Table~\ref{tab:threshold_QHW}. \tool computes $t_{m}$ by dividing $T1$ by the average \textit{gate} execution time. It then filters out those whose $n_{m}$ is greater than $t_{m}$. Take \codeIn{ibm\_16\_melbourne} as an example: with $T1$ = 53.57$\mu s$ and the average gate execution time = 928$\mathit{ns}$, $t_{m}$ is 53570/928=57. A circuit whose total number of moments is above 57 is filtered out for \codeIn{ibm\_16\_melbourne}. 

The threshold $\delta_{2qubit}$ is determined in the following way. Suppose a user is willing to tolerate an addition error rate of $t$ for the entire quantum program's final measurements (with 0.1 as the default). Using $t$, we compute $\delta_{2qubit}$ as the maximum number of 2-qubit gates to be added or deleted from the number of 2-qubit gates in the original circuit. Suppose that CNOT's error rate for this IBM computer is 1.07\%. If CNOT is used $d_{2qubit}$ times in a row {additionally}, its updated error rate would be by $1-(1-e)(1-0.0107)^{d_{2qubit}}$, where $e$ is the original error rate. Since both $e$ and $(1-0.0107)^{d_{2qubit}}$ are relatively small, we can regard the error rate change as $1-(1-0.0107)^{d_{2qubit}}$. Therefore $|d_{2qubit}-\delta_{2qubit}|$ should be less than $log_{(1-0.0107)}{(1-t)}$. $\delta_{2qubit}$ is 11 when $t$ is 0.1. 
The above thresholds and filtering condition in Definition~\ref{th_def} are customizable according to hardware's published error rates, supported gate types, and T1 relaxation time.


\begin{table}[t]\small
    \centering   \caption{Cumulative probability of KS test}\label{tab:kstest}
    
\scalebox{0.9}{
    \begin{tabular}{ccc||ccc}
    \hline
        {\bf Measurement} & \multicolumn{2}{c|}{\bf State} & {\bf Cumulative} & \multicolumn{2}{c}{\bf State}\\\cline{2-3}\cline{5-6}
        
        {\bf Distribution} & {\bf `0'} & {\bf `1'} &  
        {\bf Probability} & {\bf `0'} & {\bf `1'}\\\hline
        $\mathit{A_1}$ & 464 & 546 & $\mathit{EDF_{A1}}$ & 0.464 & 1 \\
        $\mathit{A_2}$ & 500 & 500 & $\mathit{EDF_{A2}}$ & 0.500 & 1 \\\hline
    \end{tabular}
}
    \vspace{-2em}
\end{table}

\subsection{Equivalence Checking via Distribution Comparison}\label{sec:equivalence_checking}
Nondeterministic nature of quantum programs makes it difficult for equivalence checking.
Developers usually reason about the output of a quantum circuit by executing it multiple times to obtain a distribution. While numerous distribution comparison methods are well studied in statistics, one consequent yet over-looked question for quantum computing is that {\em how many measurements do we need for a reliable evaluation to ensure the relative error between two distributions is within a given threshold $t$ with confidence $p$}? We design a novel equivalence checking component, which consists of: (1) a particular distribution comparison method $C$, and (2) an estimation of the required number of measurements  for $C$, given a threshold $t$ and confidence $p$. \tool is equipped with {\em K-S test} and {\em Cross Entropy}, but is also extensible to other comparison methods by providing a new comparison-specific measurement estimation.

\MyPara{K-S Test}~\cite{kolmogorov1933sulla} has been used to check the equality of distributions by measuring the largest vertical distance between empirical distribution functions (EDFs) in two steps. First, it creates the EDF for a given distribution by calculating the cumulative probability of different outcome states with respect to the total number of samples. In Table~\ref{tab:kstest}, $\mathit{A_1}$ and $\mathit{A_2}$ are two state distributions for 1000 samples. $\mathit{A_1}$ has 464 samples in state `0' while 546 samples in state `1'. Thus, the consequent $\mathit{EDF_{A1}}$ indicates that the cumulative probability of $A_1$ samples in state `0' and state `1' are 0.464 (464/1000) and 1 ((464+546)/1000) respectively. Next, K-S test calculates the largest distance $D$ of EDFs for each state, and uses such distance to quantify the difference of the original distributions under comparison. In Table~\ref{tab:kstest}, the largest distance is 0.06 ($|0.464-0.500|$) for state `0'.

\tool evaluates this K-S distance with a user-defined threshold $t$. If the K-S distance of two results is less than $t$, \tool regards them as similar results. \tool provides a statistical guarantee on this comparison by estimating the required number of samples. For two distributions $d1$ and $d2$ over $m$ outcome states, prior work~\cite{chan2014optimal} theoretically ensures that, with $n=\Omega(m^{1/2}\cdot t^{-2})$ samples, a sample-optimal tester can check if the relative error of $L1$ distance is within a threshold $t$ when using a default confidence level of $p$=2/3.  \tool estimates $n$ using Equation~\ref{eq:ks_number_samples}. This estimated number is directly applicable to \tool because K-S distance is bounded by $L1$ distance~\cite{kolmogorov1933sulla}. In other words, Equation~\ref{eq:ks_number_samples} calculates the number of measurements required, parameterized with respect to $p$. We empirically set $p$ as 2/3, as it is a commonly used default in quantum volume measurement~\cite{aaronson2016complexity, cross2019validating}, bioinformatics, and other statistics comparison. 

\begin{equation}
    n = A \cdot \frac{1}{\sqrt{1-p}} \cdot m^{1/2} \cdot t^{-2}
    \label{eq:ks_number_samples}
\end{equation}
\noindent where $A$ is a platform-related constant and $m$ is the number of qubit states. In Figure~\ref{fig:result_1q}, 2828 measurement samples are needed, when we empirically set $t=0.1$, $p=2/3$, and $A=12$ for Qiskit. In our evaluation, we empirically measure the constant $A$ for each platform by repeatedly running the same programs and compare the results.



\begin{figure}[t]
    \begin{subfigure}[b]{\linewidth}
    \centering
    \input{code/approach_Qsimulator}
    \caption{Quantum simulator.}
        \label{fig:1q_qvm}
    \end{subfigure}
        
    \begin{subfigure}[b]{\linewidth}
    \centering
    \input{code/approach_matrix}
    \caption{State-Vector simulator.}
        \label{fig:1q_state}
    \end{subfigure}
\vspace{-1.5em}
\caption{A quantum circuit in Qiskit with two backends: quantum simulator and state-vector simulator. \label{fig:result_1q}}
\vspace{-1em}
\end{figure}

\MyPara{Cross Entropy} measures the difference between distributions via the total entropy. It represents the average number of bits needed to encode data coming from an underlying distribution $q_{1}$ when we use an estimated target distribution $q_{2}$.

\begin{equation}
    H(q_{1},q_{2}) = -\sum_{x=0}^{max} q_{1}(x) \log q_{2}(x)
\end{equation}



Prior work ensures that, with $n=\Omega(m^{2/3}\cdot t^{-4/3})$ samples, the expected cross entropy of two similar distributions over $m$ outcome states can be bounded with $t$~\cite{chan2014optimal,shi2019accurate}. $t$ is the difference from $H(q_{1},q_{1})$. Similar to K-S test, \tool estimates the number of required samples to reliably satisfy this bound, as shown in Equation~\ref{eq:cs_numbers}. 
\begin{equation}\label{eq:cs_numbers}
    n = A \cdot \frac{1}{\sqrt{1-p}}\cdot m^{2/3} \cdot t^{-4/3}
\end{equation}
For Figure~\ref{fig:result_1q}, we need 420 measurements with $t=0.1$, $p=2/3$, and $A=7$ for Qiskit. This measurement trials are different from K-S test, because we are using different distance metrics. 

\MyPara{Comparison with Reference Distribution.}
After generating a group of equivalent programs and filtering out worthless circuits, \tool executes the remaining circuits and calculates the average distribution from their results. \tool will regard this average distribution as the {\it reference distribution}. With the distribution comparison methods (eg. K-S Test, Cross-entropy, etc), \tool compares this reference distribution with each result distribution and reports divergence when the distance is larger than the threshold $t$. 

\MyPara{Reporting Divergence Explanation.} \tool reports the potential source of the divergence in program $P$: 
\begin{enumerate}
    \item If $P$ finds divergence when using different backends while keeping a frontend's options unchanged, \tool reports this as a potential backend source;
    \item If $P$ finds divergence when using a specific backend while varying a frontend's  options, \tool reports this as a potential frontend source;
    \item Otherwise, \tool reports this as other sources, such as a potential bug in the API gate implementation.
\end{enumerate}

\section{Evaluation}
\label{sec:Eva}
We evaluate following research questions:
\begin{description}[leftmargin=3em,style=nextline]
\item[RQ1] How many syntactically different programs can be generated by \tool's mutation and equivalent gate transformation? 

\item[RQ2] How much speedup can we achieve via filtering and obviating the need of invoking a quantum simulator or hardware?

\item[RQ3] What has \tool found via differential testing of the widely-used QSSes?
\end{description}

\MyPara{Benchmarks.} 
\tool starts differential testing with five well known quantum algorithms as seed programs~\cite{grover1996fast,shor1994algorithms,bernstein1997quantum,mcclean2016theory,deutsch1992rapid} ({Deutsch-Jozsa, Berstein-Vazira, VQE--Variational-Quantum-Eigensolver, Grover, and QAOA--Quantum Approximate Optimization Algorithm}) and one additional program {X Gate} listed in Table~\ref{tab:subjects}. We do not use the relatively large algorithms like Shor's\cite{shor1994algorithms} because IBM's public access can support up to 16 qubits only. Large algorithms also require many moments and 2-qubit gates, often producing unreliable results on quantum hardware.

\begin{table*}[t]
    \centering  \caption{Seed subject programs}
    \vspace{-2ex}
    \label{tab:subjects}
    \scalebox{0.95}{
    \begin{tabular}{clllllll}
    \toprule
        {\bf ID} & {\bf Program} & {\bf \# of Qubits} &{\bf Moments}&{\bf 2-qubit gate} & {\bf Description} &{\bf Iteration Number} &{\bf Measurement trial with $t=0.1$} \\\hline
        P1 & X gate & 1 & 1 & 0 & one-qubit $X$ gate & 46 & 2000$\sim$2828\\
        P2 & Deutsch-Jozsa & 4 & 39 & 33  & check if a function is balanced & 95 &5293$\sim$7998\\
        P3 & Bernstein-Vazira & 4 & 41 & 32 & find {a} and {b} for {f(x) = ax+b} & 121 & 5293$\sim$7998\\
        P4 & Grover & 5 & 84 & 53 & find a unique input in a database & 129 & 8000$\sim$11312\\
        P5 & VQE & 4 & 36& 28 & approximate the lowest energy level & 171 &5293$\sim$7998\\
        P6 & QAOA & 5 & 29 & 19 & QAOA algorithm & 168 &8000$\sim$11312\\
    \bottomrule
    \end{tabular}
     }
     \vspace{-3ex}
\end{table*}

\MyPara{Experimental Environment.} We evaluate \tool with three widely-used QSSes: Pyquil 2.19.0 with Quilc 1.19.0, Qiksit 0.21.0, and Cirq 0.9.0. 

We run the circuits on five different hardware versions based on their availability, including \codeIn{ibmq\_santiago}, \codeIn{ibmq\_yorktown}, \codeIn{ibmq\_16\_melbourne}, \codeIn{ibmq\_belem}. and \codeIn{ibmq\_quito}. The details of hardware can be found on IBM's quantum computing website~\cite{qiskit_rpo}. We use K-S test with $t=0.1$ as the distribution comparison method.

\subsection{RQ1 Variant Generation via S2S transformation} 

\begin{figure}[t]
\centering
    \captionsetup[subfigure]{justification=justified,singlelinecheck=false}
    \begin{adjustbox}{minipage=\textwidth,scale=0.93}
    \begin{subfigure}[b]{1\textwidth}
    \scalebox{0.75}{\begin{tikzpicture}
\begin{axis}[
symbolic x coords={P1,P2,P3,P4,P5,P6},
	xtick=data,
    width=0.6\textwidth,
    height=0.27\textwidth,
    bar width = 0.25cm,
	ylabel={\# Variants},
	 legend pos=north west,
     legend columns=3,
    legend style={at={(0.02,0.8)},anchor=south west, column sep=5pt},
	ybar ,
   label style={font=\large},
	ymin=0,
	    ymax=300,
	  ylabel near ticks,
           xlabel near ticks,
           nodes near coords,
          enlarge y limits={0.3,upper},
          enlarge x limits=0.1, 
           bar shift={-\pgfplotbarwidth*1.6/2}
]
\addplot [fill=gray]
coordinates{(P1,46)(P2,95)(P3,121)(P4,129)(P5,171)(P6,168)};
\legend{\# Variants created via mutations}
\end{axis}

\begin{axis}[
symbolic x coords={P1,P2,P3,P4,P5,P6},
	xtick=data,
    width=0.6\textwidth,
    height=0.27\textwidth,
    bar width = 0.25cm,
	ylabel={\# Circuits},
	 legend pos=north west,
    legend style={at={(0.02,0.6)},anchor=south west, column sep=5pt},
	ybar ,
   label style={font=\large},
	ymin=0,
	    ymax=4000,
	  axis y line*=right,
	  ylabel near ticks,
	  axis x line=none,
	  nodes near coords,
          enlarge y limits={0.3,upper},
          enlarge x limits=0.1, 
        bar shift={\pgfplotbarwidth*1.6/2}
]

\addplot  [fill=white] 
coordinates{(P1,921)(P2,2103)(P3,2487)(P4,2021)(P5,3721)(P6,3546)};
\legend{\# Total number of circuits}
\end{axis}
\end{tikzpicture}}
    \vspace{-1ex}
    \caption{Total circuits and variants originating from each seed algorithm }
    \vspace{-0.5ex}
    \label{fig:total_circuit}
    \end{subfigure}
    \hfill
    \vspace{-1.5em}
    \begin{subfigure}[b]{1\textwidth}
    \scalebox{0.75}{\begin{tikzpicture}
\begin{axis}[
symbolic x coords={P1,P2,P3,P4,P5,P6},
	xtick=data,
    width=0.6\textwidth,
    height=0.27\textwidth,
    bar width = 0.25cm,
	ylabel={\% circuits worthwhile to run},
	 legend pos=north west,
     legend columns=3,
    legend style={at={(0.02,0.8)},anchor=south west, column sep=5pt},
	ybar ,
   label style={font=\large},
   nodes near coords={\pgfmathprintnumber\pgfplotspointmeta\%},
	ymin=0,
	    ymax=100,
	  ylabel near ticks,
	   yticklabel={$\pgfmathprintnumber{\tick}\%$},
           xlabel near ticks,
          enlarge y limits={0.3,upper},
          enlarge x limits=0.1, 
          bar shift={-\pgfplotbarwidth*1.7/2}
]

\addplot  [fill=gray] 
coordinates{(P1,19)(P2,31)(P3,28)(P4,30)(P5,37)(P6,42)};
\legend{\% Circuits worthwhile to run}
\end{axis}

\begin{axis}[
symbolic x coords={P1,P2,P3,P4,P5,P6},
	xtick=data,
    width=0.6\textwidth,
    height=0.27\textwidth,
    bar width = 0.25cm,
	ylabel={Time saving (hours)},
	 legend pos=north west,
    legend style={at={(0.02,0.6)},anchor=south west, column sep=5pt},
	ybar ,
   label style={font=\large},
	ymin=0,
	    ymax=300,
	  axis y line*=right,
	  ylabel near ticks,
	  axis x line=none,
	  nodes near coords,
          enlarge y limits={0.3,upper},
          enlarge x limits=0.1, 
        bar shift={\pgfplotbarwidth*1.7/2}
]

\addplot  [fill=white] 
coordinates{(P1,21)(P2,83)(P3,72)(P4,94)(P5,242)(P6,233)};
\legend{Time saving (hours)}
\end{axis}

\end{tikzpicture}}
    \vspace{-1ex}
    \caption{The percentage of circuits that are worth running}
    \label{fig:worthwhile}
    \end{subfigure}

    \end{adjustbox}
\caption{Statistics of \tool-generated circuits.}
    \vspace{-1em}
\end{figure}
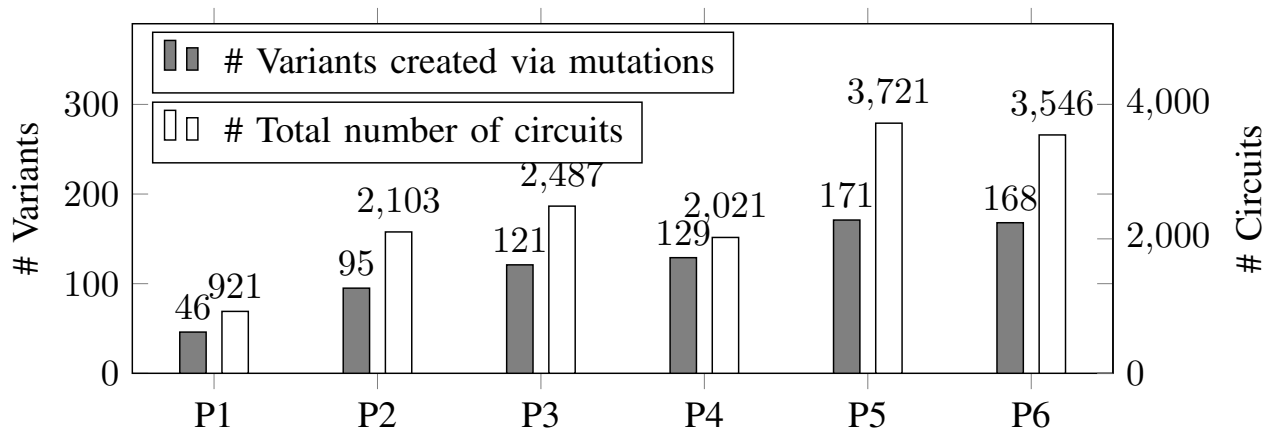
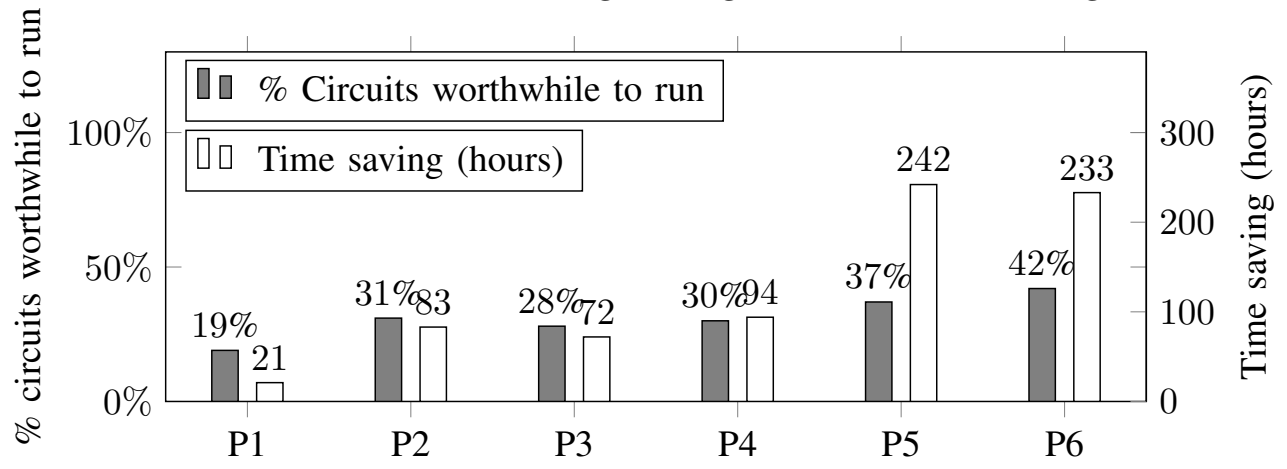

 As shown in Figure~\ref{fig:total_circuit}, for all six seed algorithms together, \tool generates 730 program variants through semantics-modifying mutations and generates the total of 14799 circuits with equivalent gate transformation to each generated variant. This total circuit generation process takes around 14 hours.
 
 Take P2 Deutsch-Jozsa algorithm as an example. 95 different program variants are generated through semantics-modifying mutations. For each variant program, \tool generates 20 logically equivalent circuits. For P2, the total generation for 2103 circuits takes around 2 hours, while the rest of differential execution via simulation or hardware execution takes around 2 days. This implies that the bottleneck of testing is not about input program generation but the execution of the generated programs, which justifies our approach to select which circuits are worthwhile to run.




\subsection{RQ2 Speed Up}

As shown in Figure~\ref{fig:worthwhile}, after filtering, only 19\%-42\% of the generated circuits are retained for differential execution on quantum hardware, leading to a 66\% reductions in quantum hardware or noisy simulator invocations. \tool finishes the entire testing process within around 17 days by leveraging its circuit selection process, which means \tool would have saved additional 30 days of testing time by filtering.


Take QAOA as an example, when running on IBM quantum hardware, the experiment would take around 7 minutes to wait in line on average (as launching a quantum job uses a shared web service for a few IBM computers in the world) and 10 seconds to execute on hardware. If users run all 3546 circuits, the total clock time would be 17 days. With filtering, \tool removes 68\% circuits that are not worthwhile to run and finishes the execution in 7 days.




\begin{figure}[t]
    \centering
    \scalebox{0.7}{\begin{tikzpicture}
\begin{axis}[
symbolic x coords={P1,P2,P3,P4,P5,P6},
	xtick=data,
    width=0.6\textwidth,
    height=0.27\textwidth,
    bar width = 0.25cm,
	ylabel={\# Generated Circuits},
	 legend pos=north west,
     legend columns=3,
    legend style={at={(0.02,0.8)},anchor=south west, column sep=5pt},
	ybar ,
   label style={font=\large},
	ymin=0,
	    ymax=2000,
	  ylabel near ticks,
           xlabel near ticks,
           nodes near coords,
          enlarge y limits={0.3,upper},
          enlarge x limits=0.1, 
           bar shift={-\pgfplotbarwidth*1.6/2}
]
\addplot [fill=gray]
coordinates{(P1,175)(P2,652)(P3,696)(P4,606)(P5,1377)(P6,1489)};
\legend{\# Total circuits}
\end{axis}

\begin{axis}[
symbolic x coords={P1,P2,P3,P4,P5,P6},
	xtick=data,
    width=0.6\textwidth,
    height=0.27\textwidth,
    bar width = 0.25cm,
	ylabel={\# Divergences on HW},
	 legend pos=north west,
    legend style={at={(0.02,0.6)},anchor=south west, column sep=5pt},
	ybar ,
   label style={font=\large},
	ymin=0,
	    ymax=40,
	  axis y line*=right,
	  ylabel near ticks,
	  nodes near coords,
	  axis x line=none,
          enlarge y limits={0.3,upper},
          enlarge x limits=0.1, 
        bar shift={\pgfplotbarwidth*1.6/2}
]

\addplot  [fill=white] 
coordinates{(P1,1)(P2,3)(P3,2)(P4,3)(P5,8)(P6,12)};
\legend{\# Circuits producing divergences}
\end{axis}
\end{tikzpicture}}
    \caption{Circuits producing divergences on IBM hardware}
    \label{fig:divergence_HW_data}
\end{figure}
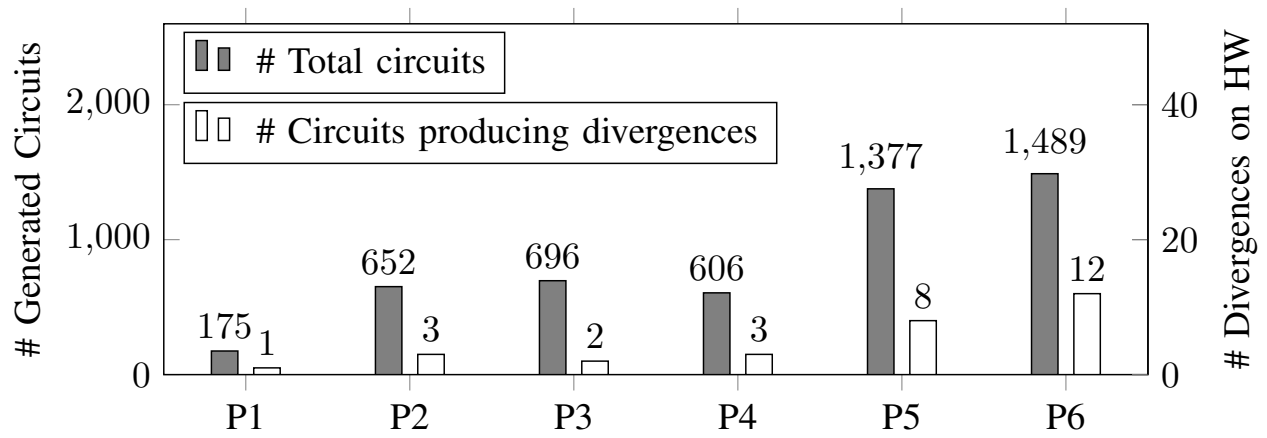

\subsection{RQ3: What has \tool found?}

\begin{table}[t]
\caption{Bugs found by \tool when executing generated programs with  simulation only}
    \centering
    \scalebox{0.87}{    \scalebox{0.95}{\begin{tabular}{c|l|l}
    \hline
        {\bf Platform}  & {\bf Bugs Description} & {\bf Source of Bugs} \\\hline
        {\bf Cirq} & Program runs endlessly with some compiler settings & Compiler Setting  \\\hline
        \multirow{3}{*}{\bf pyquil}  & Simulator register wrong number of qubits & Simulator Backends \\\cline{2-3}
         & Crashes on control gates on certain backends& Gate implementation  \\\cline{2-3}
         & Simulator stuck into a bad state & Simulator Backends \\\hline
    \end{tabular}}
}
    \label{tab:error_found}
\end{table}

From the 730 sets of semantically equivalent circuits, \tool found 33 differing outcomes out of 730. 
4 out of 33 are crashes in simulators. The remaining 29 cases are divergence beyond expected noise on IBM hardware. 
By inspecting all 33 cases carefully, we determined total 6 sources of instabilities: 4 simulator crashes and 2 potential root causes that may explain 25 out of 29 cases of divergence on IBM hardware. For the remaining 4 divergence cases, we could not easily determine their underlying root causes.


\subsubsection{Crash bugs in  simulators\label{sec:non-quantum}}

\tool reports {four} crashes during differential testing with both noiseless and noisy simulators.
All divergences involve clear failure signals. All are due to bugs in compiler or simulator implementations, summarized in Table~\ref{tab:error_found}. 2 out of 4 crashes were already reported by developers~\cite{pyquil_usedtoomanyqubits,pyquil_bad_simulator} and 2 out of 4 crashes were confirmed by developers, when we filed crash reports~\cite{cirq_qdiff_report_bug,pyquil_qdiff_report_bug1}.

\MyPara{Compiler option error:}
Given an arbitrary program in Cirq, when the compiler option \codeIn{clear\_span} is set to a negative number or \codeIn{clear\_qubits} is set to an unregistered qubit, the execution does not terminate. Cirq does not check the boundary values of compiler options and attempts to reset non-existing qubits. This bug was detected during explorations of compiler settings. When these compiler options are set to the aforementioned values, \tool notices that the execution does not finish in a reasonable time, indicating a potential infinite loop. \tool found this bug on the 74$^{th}$ iteration with P3. 

\begin{figure}[t]
        \begin{subfigure}[h]{.23\textwidth}
        \input{code/evaluation_pyquil_3}
        \caption{\codeIn{run\_and\_measure} measures all 16 qubits in device \codeIn{Aspen-0} when a simulator allocates 3 qubits, resulting in an exception.}
            \label{fig:pyquil_error_3}
        \end{subfigure}
        \hfill
        \begin{subfigure}[h]{.22\textwidth}
        \input{code/evaluation_pyquil_2}
        \caption{Controlled X gate raises an error when 2 qubits are allocated for simulation.}
            \label{fig:pyquil_error_2}
        \end{subfigure}
    \vspace{-1ex}
    \caption{Bugs in Pyquil found by \tool.
    \label{fig:pyquil_error}}
    \vspace{-2em}
\end{figure}

\MyPara{Backend registers a wrong number of qubits:} \tool found that Pyquil's measurement crashed on  \codeIn{Aspen-0-xQ-A-qvm}. In Pyquil, users can specify the quantum simulator to have the same topology as a real 16 qubit device \codeIn{Aspen-0-16Q-A} by setting the backend to be \codeIn{Aspen-0-16Q-A-qvm} (\codeIn{16Q} refers to using 16 qubits in simulation). However, \tool found that when a user allocates to use 3 qubits in simulation (line 1 in Figure~\ref{fig:pyquil_error_3}) but attempts to conduct measurements at line 2 using \codeIn{run\_and\_measure}, Pyquil simulator crashes due to a wrong number of allocated qubits. This bug was found with \tool's backend exploration. Another user reported the same issue on Pyquil's Github~\cite{pyquil_usedtoomanyqubits}. 

\MyPara{Simulator is stuck into a bad state:} 
Pyquil's simulator raises an exception when taking an empty circuit as input. Afterward, all subsequent invocations to the simulator crash even when the input circuit is valid and not empty. \tool detects this bug by mutating an input program to an empty program and executing it on both \codeIn{pyqvm} simulator and state-vector simulator. While the state-vector simulator runs this empty circuit normally, \codeIn{pyqvm} throws an exception. Then, \tool generates an arbitrary non-empty program. In the next few iterations, \tool finds this bug because \codeIn{pyqvm} crashes, while the state-vector simulator does not. 

\MyPara{Wrong type of the controlled gate:} Figure~\ref{fig:pyquil_error_2} shows a scenario where Pyquil crashed when a control $X$ gate was used on  \codeIn{pyqvm} with 2 qubit allocation. The exception message shows \textit{``ValueError: cannot reshape array of size 4 into shape (2,2,2,2)''}. A control $X$ gate should be a 2-qubit gate (which is 4 $\times$ 4 matrix), but the gate was represented as 1 $\times$ 4 by \codeIn{pyqvm}, which is a bug. This bug was found in the 15$^{th}$ iteration with P1 as a seed. 

\subsubsection{Divergences on real hardware}

\label{sec:instability}
Figure~\ref{fig:divergence_HW_data} reports the total numbers of circuits executed on quantum hardware and those that exhibit behavioral divergence. The rate of latter is relatively low (0.3\%-0.8\%), which demonstrates the robustness of the quantum software and hardware we tested—this is not surprising since such software and hardware has been widely used and continuously improved
for a number of years. On the other hand, these results also
highlight the need of a systematic testing framework such as
\tool for quantum developers—since quantum bugs are rare
and hard-to-detect, developers should test their programs/tools
exhaustively with a \tool-like approach before releasing
them.

For 29 divergence cases on IBM hardware, we manually inspected the corresponding circuits. We then determined 2 root causes that may explain 25 out of 29 cases. For the remaining 4 cases, we could not easily determine underlying root causes. We discuss each root cause with examples in this subsection.

\begin{figure}[t]
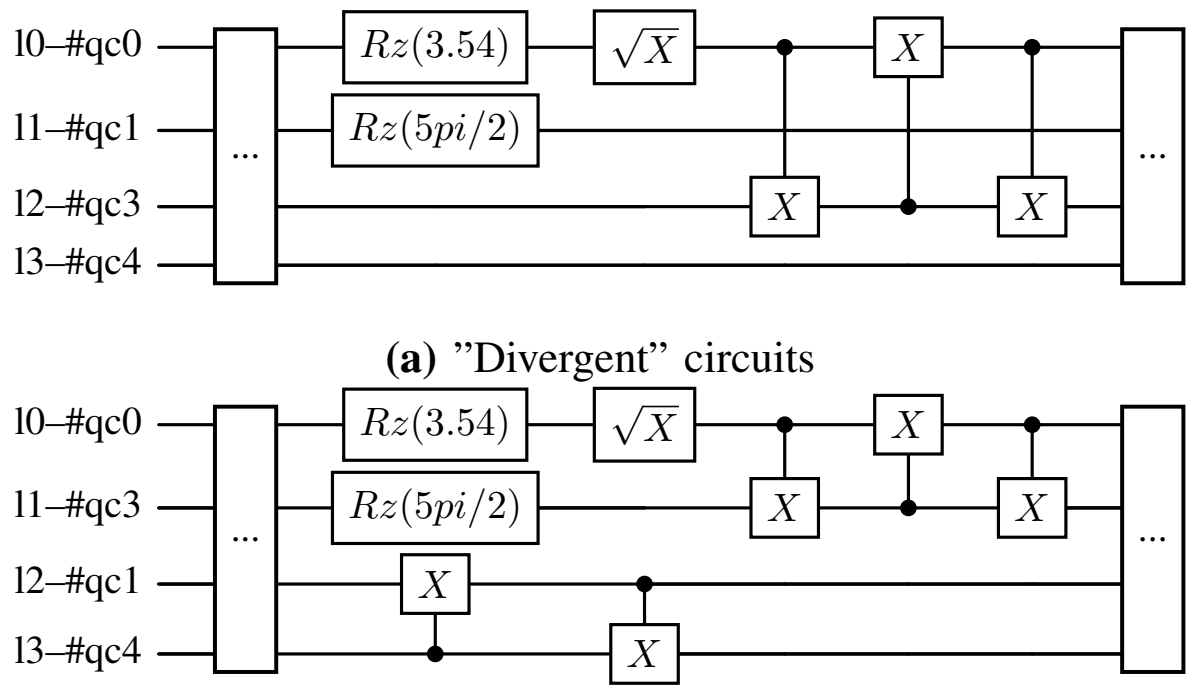

    \centering
    \begin{adjustbox}{minipage=\textwidth,scale=0.93}
    \begin{subfigure}[b]{0.5\textwidth}
    \centering
    \scalebox{0.8}{\input{code/error2_circuit1}}
    \caption{"Divergent" circuits}
    \label{fig:divergence_HW_a2}
    \end{subfigure}
    
    \begin{subfigure}[b]{0.5\textwidth}
    \centering
    \scalebox{0.8}{\input{code/error2_circuit2}}
    \caption{"Good" circuits}
    \label{fig:divergence_HW_b2}
    \end{subfigure}
    \end{adjustbox}
    \caption{Divergence on IBM \codeIn{ibmq\_yorktown}: no operation on qc4 for a long time.}
    \label{fig:divergence_HW_example2}
\end{figure}

\begin{figure}[t]
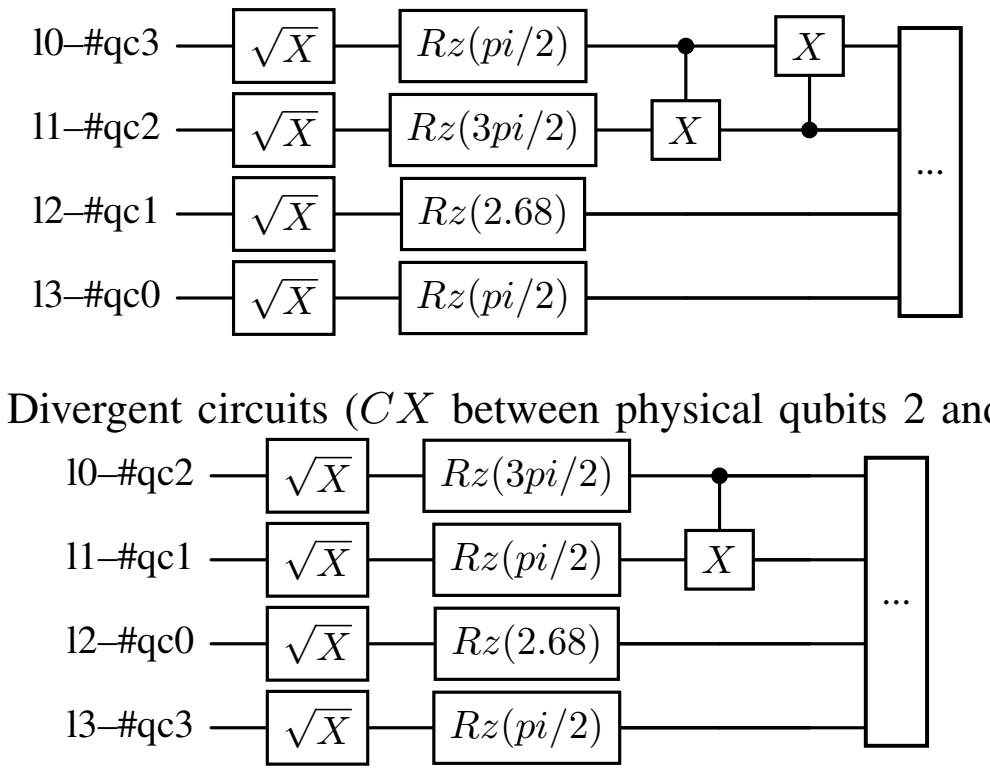

    \centering
    \begin{adjustbox}{minipage=\textwidth,scale=0.93}
    \begin{subfigure}[b]{0.5\textwidth}
    \centering
    \scalebox{0.8}{\input{code/error1_circuit1}}
    \caption{Divergent circuits ($CX$ between physical qubits 2 and 3)}
    \label{fig:divergence_HW_a}
    \end{subfigure}
    
    \begin{subfigure}[b]{0.5\textwidth}
    \centering
    \scalebox{0.8}{\input{code/error1_circuit2}}
    \caption{"Good" circuits ($CX$ between physical qubits 1 and 2)}
    \label{fig:divergence_HW_b}
    \end{subfigure}
    \end{adjustbox}
    \caption{Divergence detected on IBM hardware: bad connection between qc\_2 and qc\_3.}
    \label{fig:divergence_HW_example}
    \vspace{-2em}
\end{figure}

\MyPara{Divergence due to 2 qubit gate errors} 
We concluded that 1, 2, 6, and 7 divergences in P1, P3, P5, and P6 respectively\textemdash 55\% of all divergences detected on hardware in total\textemdash can be explained by placing many 2 qubit gates  between so called {\it couplers} qubits. IBM released the reliability of connection between each qubit pair on their hardware~\cite{qiskit_rpo}. 
For example, in hardware \codeIn{santiago}, mapping $CX$ between physical qubits \{2, 3\} is less reliable than mapping $CX$ between qubits \{1, 2\}~\cite{qiskit_rpo}. We found 16 out of 29 divergences could have the same underlying cause of using $CX$ between known, unreliable qubit connections. These errors could be reduced through improved mapping to 2-qubit gates or using other strategies such as randomized compilation~\cite{wallman2016noise,hashim2020randomized}. 

Consider the two equivalent circuits shown in Figure~\ref{fig:divergence_HW_example}, generated from P6 QAOA. These two circuits generated divergent measurements on \codeIn{santiago}, although both circuits have nearly the same moments and the same number of 2-qubit gates. This is because the $CX$ error rate of physical qubits 2 and 3 is much higher than qubits 1 and 2 (77\% higher according to published information from IBM~\cite{qiskit_rpo}). It appears that Qiskit's logical to physical qubit mapping procedure does not always avoid the use of $CX$ on qubits 2 and 3 in their compilation and qubit allocation steps. Thus, Figure~\ref{fig:divergence_HW_a} produces divergence from Figure~\ref{fig:divergence_HW_b}. 

\MyPara{Divergence due to qubit dephasing \& decoherence:} 9 of 29 divergences\textemdash 2, 3, 2, and 2 divergences in P2, P4, P5, and P6\textemdash  
could have the same underlying cause of qubit {dephasing \& decoherence}. Qubits that remain idle for long periods tend to dephase and decohere~\cite{Cirq_fix_cross_talk}. Figure~\ref{fig:divergence_HW_example2} shows a pair of circuits with a similar depth and a similar number of 2-qubit gates. However, when run on hardware \codeIn{ibm\_belem}, the pair produces divergences beyond expected noise. We speculate that, when no operation is applied to physical qubit 4 for 18 moments, it may increase dephasing and decoherence possibilities. This can by fixed by adding two successive Pauli Y gates~\cite{nielsen2002quantum} on idle qubits during the compilation phase~\cite{Cirq_fix_cross_talk}.

\MyPara{Others:} 
For other 4 divergence, we could not easily determine underlying root causes. Because the circuit moments and the number of CX gate are roughly the same with their equivalent groups,
stochastic errors in hardware can mostly be ruled out. The
problem could be low-level quantum control software bugs
that emerge from different combinations of gates, resulting in
different control / coherent errors introduced at the pulse level.

\subsection{Threats to Validity}
\MyPara{Lack of Error Correction:}
The number of divergences on quantum hardware found by \tool would depend on the reliability of hardware and its error correction capability. If it were to run on an error-corrected quantum hardware, which does not exist yet, it may report fewer divergences and it would be easier to disambiguate whether divergences are caused by software-level defects as opposed to hardware-level defects. 

Similarly, 2-qubit gate errors depend on which qubit connections that the gates are applied to, Therefore, it may be necessary to adjust the divergence threshold $t$ based on the empirical 2 qubit error rates for each connection and how many times the 2 qubit gates are used on that connection. Such impact of 2 qubit error rates must be investigated further. 

\MyPara{Time Out:} In fuzz testing, longer experimentation periods tend to expose more errors or new program execution paths~\cite{evaluatefuzzing}. The total time taken for all our experiments was limited to seven days.

\MyPara{Number of Qubits:} The maximum number of qubits that we used for our experiment was 5 qubits, because the only publicly available hardware limits public access up to 16 qubits and the waiting time tends to increase significantly, as you request more qubits. Running experiments on Google's sycamore processor with 53 qubits and 1000+ 2q gates may produce different results~\cite{mi2021information}.

\section{Related Work}
\label{sec:related}
\vspace{-.5em}
\MyPara{Compiler and Framework Testing.} Random differential testing (RDT)~\cite{mckeeman1998differential,tang2020compiler} is a widely-used technique that compiles the same input program with two or more compilers that implement the same specification. Equivalence modulo inputs (EMI)~\cite{le2014compiler} is such an example that tests compilers by generating equivalent variants. Many random program generators are used for compiler testing~\cite{chen2016empirical}.
Csmith~\cite{yang2011finding} randomly generates C programs and checks for inconsistent behaviors via differential testing. Quest~\cite{lindig2005random} focuses on argument passing and value returning, while testing with randomly generated programs. Different from Csimth-like tools, refactoring-based testing systematically modifies input programs with refactorings, as opposed to random program generation. Orion~\cite{le2014compiler} adapts EMI to test GCC and LLVM compilers. Christopher et al.~\cite{lidbury2015many} combine random differential testing and EMI-based testing to test OpenCL compilers. Orison~\cite{le2015finding} uses a guided mutation strategy for the same purpose. 

Such classical compiler testing is not directly applicable to quantum software stacks due to the three challenges: (1) how to generate variants, (2) how to test simulators and hardware together with compilers, and (3) how to interpret quantum measurements for differential testing.  

\MyPara{Quantum Testing and Verification.} 
Zhao~\cite{zhao2020quantum} introduces a quantum software life cycle and lists the challenges and opportunities we face. Ying et al.~\cite{ying2020symbolic} formally reason about quantum circuits by representing qubits and gates using matrix-valued Boolean expressions, and verify them using a combination of classical logical reasoning and complex matrix operations. Huang et al.~\cite{huang2019statistical} introduce quantum program assertions, allowing programmers to decide if a quantum state matches its expected value. They define a logic to provides $\epsilon$-robustness to characterize the possible distance between an ideal program and an erroneous one. Proq~\cite{li2020proq} is a runtime assertion framework for testing and debugging quantum programs. It transforms hardware constraints to executable versions for measurement-restricted quantum computers. QPMC~\cite{feng2015qpmc} applies classical model checking on quantum programs based on Quantum Markov Chain. Ali et al.~\cite{ali2021assessing} propose a new testing metric called quantum input output coverage, a test generation strategy, and two new test oracles for testing quantum programs. Two test oracles include {\it wrong output oracle} that checks whether a wrong output has been returned, and {\it output probability oracle} that checks whether the quantum program returns an expected output with its corresponding expected probability. However, their work targets at quantum program testing and the measurement they used might not be sufficient enough. While all these techniques find errors in \emph{quantum programs}, \tool aims to find errors in quantum software stacks. 

Verified quantum compilers guarantee gate transformation and circuit optimization is correct by construction.  CertiQ~\cite{shi2019contract} is a verified Qiskit compiler by introducing a calculus of quantum circuit equivalence to check the correctness of compiler transformation. VOQC~\cite{hietala2019verified} provides a verified optimizer for quantum circuits by adapting CompCert~\cite{leroy2016compcert} to the quantum setting. Smith and Thornton~\cite{smith2019quantum} present a compiler with built-in translation validation via QMDD equivalence checking. As discussed in Section~\ref{sec:intro}, these tools ensure correctness in quantum gate optimizations. However, \tool is a complimentary technique based on testing and its scope includes both quantum backends and frontends. 

\MyPara{Differential Testing.} Differential testing~\cite{mckeeman1998differential} has been used to test large software systems and to find bugs in various domains such as SSL/TLS~\cite{brubaker2014using,petsios2017nezha}, machine learning applications~\cite{guo2018dlfuzz}, JVM~\cite{chen2016coverage}, and clones~\cite{zhang2017automated}, etc. Mucerts~\cite{chen2018rfc} applies differential testing to check the correctness of certificate validation in SSL/TLS. It uses a stochastic sampling algorithm to drive its input generation while tracking the program coverage. DLFuzz~\cite{guo2018dlfuzz} does fuzz testing of Deep Learning systems to expose incorrect behaviors. Chen et al.~\cite{chen2016coverage} perform differential testing of JVM with input generated from Markov Chain Monte Carlo sampling with domain-specific mutations with the knowledge of Java class file formats. 


\MyPara{Mutation-Based Fuzz Testing.} 
Fuzz testing mutates the seed inputs through a \emph{fuzzer} to maximize a specific guidance metric, such as branch coverage. It has been shown to be highly effective in revealing a diverse set of bugs, including correctness bugs~\cite{babic2019fudge,park2012empirical,hfuzz}, security vulnerabilities~\cite{de2015protocol,brennan2020jvm,gan2018collafl}, and performance bugs~\cite{wen2020memlock,lemieux2018perffuzz}. For example, AFL~\cite{afl} mutates a seed input to discover previously unseen coverage profiles. MemLock~\cite{wen2020memlock} employs both coverage and memory consumption metrics to find abnormal memory behavior.

Instead of flipping several bits or bytes in each mutation, several techniques support domain-specific mutations. SDF~\cite{lin2015browser} uses seed properties to guide mutation in web-browser fuzz testing. BigFuzz~\cite{zhangbigfuzz} designs mutations for dataflow-based big data applications. \tool is similar to this line of work by adapting mutation testing to a new domain. However, different from traditional fuzzing, \tool re-invents the notion of a test oracle by re-designing a quantum measurement comparison method in a noisy, probabilistic domain. 

\section{Conclusion}
\label{sec:conclusion}
Quantum computing has emerged to be a promising computing paradigm with remarkable advantages over classical computing. 
\tool is the first to reinvent differential testing for quantum software stacks. It adapts the notion of equivalence checking to the quantum domain,  redesigns underlying program generation and mutation methods, and optimizes differential testing to reduce compute-intensive simulation or expensive hardware invocation. It is effective in generating variants, reduce 66\% unnecessary quantum hardware or noisy simulator invocations, and uses divergence to isolate errors in both the higher and lower levels of the quantum software stack.

\section{Acknowledgment}
\label{sec:acknowledgment}

We would like to thank the anonymous reviewers for their valuable feedback. We thank Alan Ho for his kind help in selective invocation and distribution comparison, and for proofreading the draft. This work is supported in part by National Science Foundations via grants CNS-1703598, CNS-1763172, CNS-1907352, CNS-2007737, CNS-2006437, CNS-2106838, CNS-2128653, CHS-1956322 CCF-1764077, CCF-1723773, ONR grant N00014-18-1-2037, and Intel CAPA grant.

\bibliographystyle{IEEEtran}
\bibliography{IEEEabrv,sample-base}

\begin{thebibliography}{10}
\providecommand{\url}[1]{#1}
\csname url@samestyle\endcsname
\providecommand{\newblock}{\relax}
\providecommand{\bibinfo}[2]{#2}
\providecommand{\BIBentrySTDinterwordspacing}{\spaceskip=0pt\relax}
\providecommand{\BIBentryALTinterwordstretchfactor}{4}
\providecommand{\BIBentryALTinterwordspacing}{\spaceskip=\fontdimen2\font plus
\BIBentryALTinterwordstretchfactor\fontdimen3\font minus
  \fontdimen4\font\relax}
\providecommand{\BIBforeignlanguage}[2]{{%
\expandafter\ifx\csname l@#1\endcsname\relax
\typeout{** WARNING: IEEEtran.bst: No hyphenation pattern has been}%
\typeout{** loaded for the language `#1'. Using the pattern for}%
\typeout{** the default language instead.}%
\else
\language=\csname l@#1\endcsname
\fi
#2}}
\providecommand{\BIBdecl}{\relax}
\BIBdecl

\bibitem{Cirq}
\BIBentryALTinterwordspacing
Q.~A. team and collaborators, ``Cirq,'' Oct. 2020. [Online]. Available:
  \url{https://doi.org/10.5281/zenodo.4062499}
\BIBentrySTDinterwordspacing

\bibitem{pyquil_paper}
R.~S. Smith, M.~J. Curtis, and W.~J. Zeng, ``A practical quantum instruction
  set architecture,'' \emph{arXiv preprint arXiv:1608.03355}, 2016.

\bibitem{qiskit_rpo}
G.~Aleksandrowicz, T.~Alexander, P.~Barkoutsos, L.~Bello, Y.~Ben-Haim,
  D.~Bucher, F.~Cabrera-Hern{\'a}ndez, J.~Carballo-Franquis, A.~Chen, C.~Chen
  \emph{et~al.}, ``Qiskit: An open-source framework for quantum computing,''
  \emph{Accessed on: Mar}, vol.~16, 2019.

\bibitem{aleksandrowicz2019qiskit}
------, ``Qiskit: An open-source framework for quantum computing,''
  \emph{Accessed on: Mar}, vol.~16, 2019.

\bibitem{pyquil_issue}
\url{https://github.com/rigetti/pyquil/issues}, 2020.

\bibitem{Cirq_issue}
\url{https://github.com/quantumlib/Cirq/issues}, 2020.

\bibitem{stackoverflow}
\url{https://stackoverflow.com/search?q=quantum++error}, April,20 2021.

\bibitem{amy2014polynomial}
M.~Amy, D.~Maslov, and M.~Mosca, ``Polynomial-time t-depth optimization of
  clifford+ t circuits via matroid partitioning,'' \emph{IEEE Transactions on
  Computer-Aided Design of Integrated Circuits and Systems}, vol.~33, no.~10,
  pp. 1476--1489, 2014.

\bibitem{nam2018automated}
Y.~Nam, N.~J. Ross, Y.~Su, A.~M. Childs, and D.~Maslov, ``Automated
  optimization of large quantum circuits with continuous parameters,''
  \emph{npj Quantum Information}, vol.~4, no.~1, pp. 1--12, 2018.

\bibitem{le2014compiler}
V.~Le, M.~Afshari, and Z.~Su, ``Compiler validation via equivalence modulo
  inputs,'' \emph{ACM SIGPLAN Notices}, vol.~49, no.~6, pp. 216--226, 2014.

\bibitem{hietala2019verified}
K.~Hietala, R.~Rand, S.-H. Hung, X.~Wu, and M.~Hicks, ``A verified optimizer
  for quantum circuits,'' \emph{Proceedings of the ACM on Programming
  Languages}, vol.~5, no. POPL, pp. 1--29, 2021.

\bibitem{shi2019contract}
Y.~Shi, X.~Li, R.~Tao, A.~Javadi-Abhari, A.~W. Cross, F.~T. Chong, and R.~Gu,
  ``Contract-based verification of a realistic quantum compiler,'' \emph{arXiv
  preprint arXiv:1908.08963}, 2019.

\bibitem{pyquil_error}
\url{https://github.com/rigetti/pyquil/issues/1034}, 2020.

\bibitem{apiserror}
\url{https://github.com/rigetti/pyquil/issues/1002}, 2020.

\bibitem{Cirq_error}
\url{https://github.com/quantumlib/Cirq/issues/2553}, 2020.

\bibitem{Dwave}
\url{https://www.cio.com/article/3161031/d-waves-quantum-computer-runs-a-staggering-2000-qubits.html},
  2017.

\bibitem{chan2014optimal}
S.-O. Chan, I.~Diakonikolas, P.~Valiant, and G.~Valiant, ``Optimal algorithms
  for testing closeness of discrete distributions,'' in \emph{Proceedings of
  the twenty-fifth annual ACM-SIAM symposium on Discrete algorithms}.\hskip 1em
  plus 0.5em minus 0.4em\relax SIAM, 2014, pp. 1193--1203.

\bibitem{kolmogorov1933sulla}
A.~Kolmogorov, ``Sulla determinazione empirica di una lgge di distribuzione,''
  \emph{Inst. Ital. Attuari, Giorn.}, vol.~4, pp. 83--91, 1933.

\bibitem{smirnov1939estimation}
N.~V. Smirnov, ``On the estimation of the discrepancy between empirical curves
  of distribution for two independent samples,'' \emph{Bull. Math. Univ.
  Moscou}, vol.~2, no.~2, pp. 3--14, 1939.

\bibitem{good1956some}
I.~Good, ``Some terminology and notation in information theory,''
  \emph{Proceedings of the IEE-Part C: Monographs}, vol. 103, no.~3, pp.
  200--204, 1956.

\bibitem{shore1980axiomatic}
J.~Shore and R.~Johnson, ``Axiomatic derivation of the principle of maximum
  entropy and the principle of minimum cross-entropy,'' \emph{IEEE Transactions
  on information theory}, vol.~26, no.~1, pp. 26--37, 1980.

\bibitem{investigate_API_error}
\url{https://github.com/quantumlib/Cirq/issues/673}, 2020.

\bibitem{investigate_backend_error}
\url{https://github.com/quantumlib/Cirq/issues/2240}, 2019.

\bibitem{investigate_gate_error}
\url{https://github.com/quantumlib/Cirq/issues/1713}, 2019.

\bibitem{grover1996fast}
L.~K. Grover, ``A fast quantum mechanical algorithm for database search,'' in
  \emph{Proceedings of the twenty-eighth annual ACM symposium on Theory of
  computing}, 1996, pp. 212--219.

\bibitem{shor1994algorithms}
P.~W. Shor, ``Algorithms for quantum computation: discrete logarithms and
  factoring,'' in \emph{Proceedings 35th annual symposium on foundations of
  computer science}.\hskip 1em plus 0.5em minus 0.4em\relax Ieee, 1994, pp.
  124--134.

\bibitem{larose2019overview}
R.~LaRose, ``Overview and comparison of gate level quantum software
  platforms,'' \emph{Quantum}, vol.~3, p. 130, 2019.

\bibitem{Cirq_rpo}
\BIBentryALTinterwordspacing
(2019) Cirq: A python framework for creating, editing, and invoking noisy
  intermediate scale quantum (nisq) circuits. [Online]. Available:
  \url{https:/github.com/quantumlib/Cirq}
\BIBentrySTDinterwordspacing

\bibitem{pyquil_rpo}
R.~Computing, ``Pyquil documentation,'' \emph{URL http://pyquil. readthedocs.
  io/en/latest}, 2019.

\bibitem{nielsen2002quantum}
M.~A. Nielsen and I.~Chuang, ``Quantum computation and quantum information,''
  2002.

\bibitem{cirq_compilerbugs}
\url{https://quantumcomputing.stackexchange.com/questions/8694/is-there-a-mistake-in-the-vqe-ansatz-in-cirqs-tutorial},
  2019.

\bibitem{afl}
``American fuzz loop,'' \url{http://lcamtuf.coredump.cx/afl/}, 2020.

\bibitem{lidbury2015many}
C.~Lidbury, A.~Lascu, N.~Chong, and A.~F. Donaldson, ``Many-core compiler
  fuzzing,'' \emph{ACM SIGPLAN Notices}, vol.~50, no.~6, pp. 65--76, 2015.

\bibitem{mendiluzemuskit}
E.~Mendiluze, S.~Ali, P.~Arcaini, and T.~Yue, ``Muskit: A mutation analysis
  tool for quantum software testing.''

\bibitem{ali2021assessing}
S.~Ali, P.~Arcaini, X.~Wang, and T.~Yue, ``Assessing the effectiveness of input
  and output coverage criteria for testing quantum programs,'' in \emph{2021
  14th IEEE Conference on Software Testing, Verification and Validation
  (ICST)}.\hskip 1em plus 0.5em minus 0.4em\relax IEEE, 2021, pp. 13--23.

\bibitem{tannu2019not}
S.~S. Tannu and M.~K. Qureshi, ``Not all qubits are created equal: a case for
  variability-aware policies for nisq-era quantum computers,'' in
  \emph{Proceedings of the Twenty-Fourth International Conference on
  Architectural Support for Programming Languages and Operating Systems}, 2019,
  pp. 987--999.

\bibitem{aaronson2016complexity}
S.~Aaronson and L.~Chen, ``Complexity-theoretic foundations of quantum
  supremacy experiments,'' \emph{arXiv preprint arXiv:1612.05903}, 2016.

\bibitem{cross2019validating}
A.~W. Cross, L.~S. Bishop, S.~Sheldon, P.~D. Nation, and J.~M. Gambetta,
  ``Validating quantum computers using randomized model circuits,''
  \emph{Physical Review A}, vol. 100, no.~3, p. 032328, 2019.

\bibitem{shi2019accurate}
Y.~Shi, M.~Wang, W.~Shi, J.-H. Lee, H.~Kang, and H.~Jiang, ``Accurate and
  efficient estimation of small p-values with the cross-entropy method:
  applications in genomic data analysis,'' \emph{Bioinformatics}, vol.~35,
  no.~14, pp. 2441--2448, 2019.

\bibitem{bernstein1997quantum}
E.~Bernstein and U.~Vazirani, ``Quantum complexity theory,'' \emph{SIAM Journal
  on computing}, vol.~26, no.~5, pp. 1411--1473, 1997.

\bibitem{mcclean2016theory}
J.~R. McClean, J.~Romero, R.~Babbush, and A.~Aspuru-Guzik, ``The theory of
  variational hybrid quantum-classical algorithms,'' \emph{New Journal of
  Physics}, vol.~18, no.~2, p. 023023, 2016.

\bibitem{deutsch1992rapid}
D.~Deutsch and R.~Jozsa, ``Rapid solution of problems by quantum computation,''
  \emph{Proceedings of the Royal Society of London. Series A: Mathematical and
  Physical Sciences}, vol. 439, no. 1907, pp. 553--558, 1992.

\bibitem{pyquil_usedtoomanyqubits}
\url{https://github.com/rigetti/pyquil/issues/1050}, 2019.

\bibitem{pyquil_bad_simulator}
\url{https://github.com/rigetti/pyquil/issues/1034}, Nov,20 2020.

\bibitem{cirq_qdiff_report_bug}
\url{https://github.com/quantumlib/Cirq/issues/3907}, Nov,20 2020.

\bibitem{pyquil_qdiff_report_bug1}
\url{https://github.com/rigetti/pyquil/issues/1259}, Nov,20 2020.

\bibitem{wallman2016noise}
J.~J. Wallman and J.~Emerson, ``Noise tailoring for scalable quantum
  computation via randomized compiling,'' \emph{Physical Review A}, vol.~94,
  no.~5, p. 052325, 2016.

\bibitem{hashim2020randomized}
A.~Hashim, R.~K. Naik, A.~Morvan, J.-L. Ville, B.~Mitchell, J.~M. Kreikebaum,
  M.~Davis, E.~Smith, C.~Iancu, K.~P. O'Brien \emph{et~al.}, ``Randomized
  compiling for scalable quantum computing on a noisy superconducting quantum
  processor,'' \emph{arXiv preprint arXiv:2010.00215}, 2020.

\bibitem{Cirq_fix_cross_talk}
\url{https://quantumai.google/cirq/google/best\_practices\#keep\_qubits\_busy},
  2021.

\bibitem{evaluatefuzzing}
\BIBentryALTinterwordspacing
G.~Klees, A.~Ruef, B.~Cooper, S.~Wei, and M.~Hicks, ``Evaluating fuzz
  testing,'' in \emph{Proceedings of the 2018 ACM SIGSAC Conference on Computer
  and Communications Security}, ser. CCS '18.\hskip 1em plus 0.5em minus
  0.4em\relax New York, NY, USA: Association for Computing Machinery, 2018, p.
  2123–2138. [Online]. Available:
  \url{https://doi.org/10.1145/3243734.3243804}
\BIBentrySTDinterwordspacing

\bibitem{mi2021information}
X.~Mi, P.~Roushan, C.~Quintana, S.~Mandra, J.~Marshall, C.~Neill, F.~Arute,
  K.~Arya, J.~Atalaya, R.~Babbush, J.~C. Bardin, R.~Barends, A.~Bengtsson,
  S.~Boixo, A.~Bourassa, M.~Broughton, B.~B. Buckley, D.~A. Buell, B.~Burkett,
  N.~Bushnell, Z.~Chen, B.~Chiaro, R.~Collins, W.~Courtney, S.~Demura, A.~R.
  Derk, A.~Dunsworth, D.~Eppens, C.~Erickson, E.~Farhi, A.~G. Fowler, B.~Foxen,
  C.~Gidney, M.~Giustina, J.~A. Gross, M.~P. Harrigan, S.~D. Harrington,
  J.~Hilton, A.~Ho, S.~Hong, T.~Huang, W.~J. Huggins, L.~B. Ioffe, S.~V.
  Isakov, E.~Jeffrey, Z.~Jiang, C.~Jones, D.~Kafri, J.~Kelly, S.~Kim,
  A.~Kitaev, P.~V. Klimov, A.~N. Korotkov, F.~Kostritsa, D.~Landhuis,
  P.~Laptev, E.~Lucero, O.~Martin, J.~R. McClean, T.~McCourt, M.~McEwen,
  A.~Megrant, K.~C. Miao, M.~Mohseni, W.~Mruczkiewicz, J.~Mutus, O.~Naaman,
  M.~Neeley, M.~Newman, M.~Y. Niu, T.~E. O'Brien, A.~Opremcak, E.~Ostby,
  B.~Pato, A.~Petukhov, N.~Redd, N.~C. Rubin, D.~Sank, K.~J. Satzinger,
  V.~Shvarts, D.~Strain, M.~Szalay, M.~D. Trevithick, B.~Villalonga, T.~White,
  Z.~J. Yao, P.~Yeh, A.~Zalcman, H.~Neven, I.~Aleiner, K.~Kechedzhi,
  V.~Smelyanskiy, and Y.~Chen, ``Information scrambling in computationally
  complex quantum circuits,'' 2021.

\bibitem{mckeeman1998differential}
W.~M. McKeeman, ``Differential testing for software,'' \emph{Digital Technical
  Journal}, vol.~10, no.~1, pp. 100--107, 1998.

\bibitem{tang2020compiler}
Y.~Tang, Z.~Ren, W.~Kong, and H.~Jiang, ``Compiler testing: a systematic
  literature analysis,'' \emph{Frontiers of Computer Science}, vol.~14, no.~1,
  pp. 1--20, 2020.

\bibitem{chen2016empirical}
J.~Chen, W.~Hu, D.~Hao, Y.~Xiong, H.~Zhang, L.~Zhang, and B.~Xie, ``An
  empirical comparison of compiler testing techniques,'' in \emph{Proceedings
  of the 38th International Conference on Software Engineering}, 2016, pp.
  180--190.

\bibitem{yang2011finding}
X.~Yang, Y.~Chen, E.~Eide, and J.~Regehr, ``Finding and understanding bugs in c
  compilers,'' in \emph{Proceedings of the 32nd ACM SIGPLAN conference on
  Programming language design and implementation}, 2011, pp. 283--294.

\bibitem{lindig2005random}
C.~Lindig, ``Random testing of c calling conventions,'' in \emph{Proceedings of
  the sixth international symposium on Automated analysis-driven debugging},
  2005, pp. 3--12.

\bibitem{le2015finding}
V.~Le, C.~Sun, and Z.~Su, ``Finding deep compiler bugs via guided stochastic
  program mutation,'' \emph{ACM SIGPLAN Notices}, vol.~50, no.~10, pp.
  386--399, 2015.

\bibitem{zhao2020quantum}
J.~Zhao, ``Quantum software engineering: Landscapes and horizons,'' \emph{arXiv
  preprint arXiv:2007.07047}, 2020.

\bibitem{ying2020symbolic}
M.~Ying and Z.~Ji, ``Symbolic verification of quantum circuits,'' \emph{arXiv
  preprint arXiv:2010.03032}, 2020.

\bibitem{huang2019statistical}
Y.~Huang and M.~Martonosi, ``Statistical assertions for validating patterns and
  finding bugs in quantum programs,'' in \emph{Proceedings of the 46th
  International Symposium on Computer Architecture}, 2019, pp. 541--553.

\bibitem{li2020proq}
G.~Li, L.~Zhou, N.~Yu, Y.~Ding, M.~Ying, and Y.~Xie, ``Proq: Projection-based
  runtime assertions for debugging on a quantum computer,'' 2020.

\bibitem{feng2015qpmc}
Y.~Feng, E.~M. Hahn, A.~Turrini, and L.~Zhang, ``Qpmc: A model checker for
  quantum programs and protocols,'' in \emph{International Symposium on Formal
  Methods}.\hskip 1em plus 0.5em minus 0.4em\relax Springer, 2015, pp.
  265--272.

\bibitem{leroy2016compcert}
X.~Leroy, S.~Blazy, D.~K{\"a}stner, B.~Schommer, M.~Pister, and C.~Ferdinand,
  ``Compcert-a formally verified optimizing compiler,'' in \emph{ERTS 2016:
  Embedded Real Time Software and Systems, 8th European Congress}, 2016.

\bibitem{smith2019quantum}
K.~N. Smith and M.~A. Thornton, ``A quantum computational compiler and design
  tool for technology-specific targets,'' in \emph{Proceedings of the 46th
  International Symposium on Computer Architecture}, 2019, pp. 579--588.

\bibitem{brubaker2014using}
C.~Brubaker, S.~Jana, B.~Ray, S.~Khurshid, and V.~Shmatikov, ``Using
  frankencerts for automated adversarial testing of certificate validation in
  ssl/tls implementations,'' in \emph{2014 IEEE Symposium on Security and
  Privacy}.\hskip 1em plus 0.5em minus 0.4em\relax IEEE, 2014, pp. 114--129.

\bibitem{petsios2017nezha}
T.~Petsios, A.~Tang, S.~Stolfo, A.~D. Keromytis, and S.~Jana, ``Nezha:
  Efficient domain-independent differential testing,'' in \emph{2017 IEEE
  Symposium on Security and Privacy (SP)}.\hskip 1em plus 0.5em minus
  0.4em\relax IEEE, 2017, pp. 615--632.

\bibitem{guo2018dlfuzz}
J.~Guo, Y.~Jiang, Y.~Zhao, Q.~Chen, and J.~Sun, ``Dlfuzz: Differential fuzzing
  testing of deep learning systems,'' in \emph{Proceedings of the 2018 26th ACM
  Joint Meeting on European Software Engineering Conference and Symposium on
  the Foundations of Software Engineering}, 2018, pp. 739--743.

\bibitem{chen2016coverage}
Y.~Chen, T.~Su, C.~Sun, Z.~Su, and J.~Zhao, ``Coverage-directed differential
  testing of jvm implementations,'' in \emph{proceedings of the 37th ACM
  SIGPLAN Conference on Programming Language Design and Implementation}, 2016,
  pp. 85--99.

\bibitem{zhang2017automated}
T.~Zhang and M.~Kim, ``Automated transplantation and differential testing for
  clones,'' in \emph{2017 IEEE/ACM 39th International Conference on Software
  Engineering (ICSE)}.\hskip 1em plus 0.5em minus 0.4em\relax IEEE, 2017, pp.
  665--676.

\bibitem{chen2018rfc}
C.~Chen, C.~Tian, Z.~Duan, and L.~Zhao, ``Rfc-directed differential testing of
  certificate validation in ssl/tls implementations,'' in \emph{2018 IEEE/ACM
  40th International Conference on Software Engineering (ICSE)}.\hskip 1em plus
  0.5em minus 0.4em\relax IEEE, 2018, pp. 859--870.

\bibitem{babic2019fudge}
D.~Babi{\'c}, S.~Bucur, Y.~Chen, F.~Ivan{\v{c}}i{\'c}, T.~King, M.~Kusano,
  C.~Lemieux, L.~Szekeres, and W.~Wang, ``Fudge: fuzz driver generation at
  scale,'' in \emph{Proceedings of the 2019 27th ACM Joint Meeting on European
  Software Engineering Conference and Symposium on the Foundations of Software
  Engineering}, 2019, pp. 975--985.

\bibitem{park2012empirical}
J.~Park, M.~Kim, B.~Ray, and D.-H. Bae, ``An empirical study of supplementary
  bug fixes,'' in \emph{2012 9th IEEE Working Conference on Mining Software
  Repositories (MSR)}.\hskip 1em plus 0.5em minus 0.4em\relax IEEE, 2012, pp.
  40--49.

\bibitem{hfuzz}
Q.~Zhang, J.~Wang, and M.~Kim, ``Heterofuzz: Fuzz testing to detect platform
  dependent divergence for heterogeneous applications,'' in \emph{Proceedings
  of the 29th ACM Joint Meeting on European Software Engineering Conference and
  Symposium on the Foundations of Software Engineering}, ser. ESEC/FSE 2021,
  2021, p. 242–254.

\bibitem{de2015protocol}
J.~De~Ruiter and E.~Poll, ``Protocol state fuzzing of $\{$TLS$\}$
  implementations,'' in \emph{24th $\{$USENIX$\}$ Security Symposium
  ($\{$USENIX$\}$ Security 15)}, 2015, pp. 193--206.

\bibitem{brennan2020jvm}
T.~Brennan, S.~Saha, and T.~Bultan, ``Jvm fuzzing for jit-induced side-channel
  detection,'' in \emph{Proceedings of the ACM/IEEE 42nd International
  Conference on Software Engineering}, 2020, pp. 1011--1023.

\bibitem{gan2018collafl}
S.~Gan, C.~Zhang, X.~Qin, X.~Tu, K.~Li, Z.~Pei, and Z.~Chen, ``Collafl: Path
  sensitive fuzzing,'' in \emph{2018 IEEE Symposium on Security and Privacy
  (SP)}.\hskip 1em plus 0.5em minus 0.4em\relax IEEE, 2018, pp. 679--696.

\bibitem{wen2020memlock}
C.~Wen, H.~Wang, Y.~Li, S.~Qin, Y.~Liu, Z.~Xu, H.~Chen, X.~Xie, G.~Pu, and
  T.~Liu, ``Memlock: Memory usage guided fuzzing,'' in \emph{42nd International
  Conference on Software Engineering}.\hskip 1em plus 0.5em minus 0.4em\relax
  ACM, 2020.

\bibitem{lemieux2018perffuzz}
C.~Lemieux, R.~Padhye, K.~Sen, and D.~Song, ``Perffuzz: Automatically
  generating pathological inputs,'' in \emph{Proceedings of the 27th ACM
  SIGSOFT International Symposium on Software Testing and Analysis}, 2018, pp.
  254--265.

\bibitem{lin2015browser}
Y.-D. Lin, F.-Z. Liao, S.-K. Huang, and Y.-C. Lai, ``Browser fuzzing by
  scheduled mutation and generation of document object models,'' in \emph{2015
  International Carnahan Conference on Security Technology (ICCST)}.\hskip 1em
  plus 0.5em minus 0.4em\relax IEEE, 2015, pp. 1--6.

\bibitem{zhangbigfuzz}
Q.~Zhang, J.~Wang, M.~A. Gulzar, R.~Padhye, and M.~Kim, ``Bigfuzz: Efficient
  fuzz testing for data analytics using framework abstraction,'' in \emph{2020
  35th IEEE/ACM International Conference on Automated Software Engineering
  (ASE)}, 2020, pp. 722--733.

\end{thebibliography}

\end{document}